\documentclass[11pt,a4paper]{article}
\pdfoutput=1
\usepackage{myjheppub}
\usepackage{latexsym,amsfonts,amsmath,amssymb,textgreek,upgreek}
\usepackage{amsthm}
\usepackage{mathtools}
\usepackage{bbm}
\usepackage{graphicx}
\usepackage{color}
\usepackage{amsfonts}
\usepackage{caption}
\usepackage{subcaption}
\usepackage[normalem]{ulem}
\usepackage{comment}
\usepackage{url}
\usepackage{slashed}
\usepackage{bm}
\usepackage{hhline}
\usepackage{physics}
\usepackage{booktabs}
\usepackage{siunitx}
\usepackage{tikz}
\usepackage{tabu}

\usepackage{footmisc}

\usepackage{tikz}
\usetikzlibrary{decorations.pathreplacing, decorations.markings,calc,shapes.misc,decorations.pathmorphing,patterns.meta, math}
\usetikzlibrary{arrows}

\newcommand{\CG}{\mathcal{G}}
\newcommand{\CK}{\mathcal{K}}

\newcommand{\CN}{\mathcal{N}}

\newcommand{\CY}{\mathcal{Y}}

\newcommand{\IR}{\mathbb{R}}

\newcommand{\SU}{\mathrm{SU}}
\newcommand{\SO}{\mathrm{SO}}

\newcommand{\U}{\mathrm{U}}

\newcommand{\rme}{e}
\newcommand{\ii}{i}

\newcommand\be{\begin{equation}}
\newcommand\ee{\end{equation}}
\newcommand\bea{\begin{eqnarray}}
\newcommand\eea{\end{eqnarray}}

\newcommand{\nn}{\nonumber}
\renewcommand{\dd}{\mathrm{d}}

\renewcommand{\=}{\;= \;}

\renewcommand{\d}{\mathrm{d}}

\renewcommand{\nn}{\nonumber\\}
\newcommand{\wh}{\widehat}
\newcommand{\wt}{\widetilde}

\newcommand{\p}{\partial}

\newcommand{\ndt}{\noindent}

\renewcommand{\i}{i}

\newcommand{\half}{\frac12}

\newcommand{\Hv}{H}

\newcommand{\OO}{\mathrm{O}}

\newcommand{\nv}{n_\text{v}}

\newcommand{\xvec}{\textbf{x}}

\newcommand{\intprod}[2]{\langle \, #1 \,,\, #2\, \rangle}

\newcommand{\beq}{\begin{equation}}
\newcommand{\eeq}{\end{equation}}
\newcommand{\ea}{\end{eqnarray}}
\newcommand{\barr}{\begin{array}}
\newcommand{\earr}{\end{array}}

\title{Gravitational index of the heterotic string
\vspace{-0.5cm}}

\author{Yiming Chen$^1$, Sameer Murthy$^{2,3}$, Gustavo J. Turiaci$^{4}$}

\affiliation{
{${}^1$ Stanford Institute for Theoretical Physics, Stanford University, Stanford, CA 94305, USA}}

\affiliation{${}^2$ Department of Mathematics, King’s College London, The Strand, London WC2R 2LS, UK}
   
\affiliation{${}^3$  School of Natural Sciences, Institute for Advanced Study, Princeton, NJ, USA}

\affiliation{${}^4$ Physics Department, University of Washington, Seattle, WA, USA}

\abstract{
The fundamental heterotic string has a tower of BPS states whose 
supersymmetric index has an exponential growth in the charges. 
We construct the saddle-point of the gravitational path integral corresponding to this index. 
The saddle-point configuration is a supersymmetric rotating non-extremal Euclidean black hole. 
This configuration is singular in the two-derivative theory. We show that 
the addition of higher-derivative terms in four-dimensional~$\CN=2$ supergravity resolves the singularity.
In doing so, we extend the recently-developed ``new attractor mechanism" to include the effect of higher-derivative terms. 
Remarkably, the one-loop, four-derivative F-term contribution to the prepotential leads to a precise match of the gravitational and microscopic index. 
We also comment, using the effective  theory near the horizon, on the possibility of a string-size near-extremal black hole. 
Our results clarify the meaning of different descriptions of this system in the literature. 
The thermal state transitions to a winding condensate and a gas of strings without ever reaching a small black hole, 
while the index is captured by the rotating Euclidean black hole solution and is constant and thus smoothly connected to the microscopic ensemble.
}

\begin{document}

\maketitle 

\section{Introduction}

Consider a fundamental heterotic string of energy~$M$,  
with~$w$ units of winding and~$n$ units of momentum around a circle of 
finite size.
In the weakly-coupled theory the string has a tower of excited states 
whose degeneracy grows exponentially in the excitation numbers. 
In the strongly-coupled theory, the effective description of the  
string is as a source for the metric and other low-energy 
fields of supergravity, and 
the solution of the low-energy theory is generically a black hole in the~$(d+1)$-dimensional non-compact spacetime whose entropy corresponds to the 
exponentially growing degeneracy of states~\cite{Bowick:1985af,Susskind:1993ws,Horowitz:1996nw}. 

When all the right-movers are in their ground states, the heterotic string
configuration preserves half the supersymmetry of the theory and 
obeys~$M^2=Q_R^2/2$ 
where~$Q_R$ is the right-moving momentum. 
The resulting tower of $\frac12$-BPS Dabholkar-Harvey states has a degeneracy which grows 
as~$\exp(4 \pi \sqrt{nw})$~\cite{Dabholkar:1989jt}. 
What is the strongly-coupled description of this ensemble of supersymmetric states? 
One natural possibility is that they form a supersymmetric black hole. Such a black hole
would be extremal with~AdS$_2 \times S^{d-1}$ near-horizon geometry. In this paper we mainly discuss four non-compact directions ($d=3$), 
although we make a few comments on other dimensions as well.  

As is well-known, this simple picture runs into various problems.\footnote{Notably, this picture does hold in three- and four-charge 
systems, and leads to the famous agreement of statistical and thermodynamic entropy of black holes in string theory~\cite{Strominger:1996sh}.} 
The zeroth-order problem, which was already discussed in the early days,  
is that the two-derivative supergravity solutions with a fundamental string source 
is singular, in that the horizon lies on top of the curvature singularity~\cite{Dabholkar:1995nc}. 
A more sophisticated analysis suggested that including higher-derivative terms 
in the low-energy supergravity stretches the horizon to string scale, thus resolving the singularity~\cite{Sen:1995in}, \cite{Sen:2004dp,Hubeny:2004ji,Dabholkar:2004dq}.

However, there are still problems with this picture. 
Firstly, if we think of obtaining the extremal black hole solution
% by lowering the temperature to zero, 
by lowering the mass to its extremal value, 
there is good evidence  that there is 
a transition at low temperatures to the so-called Horowitz-Polchinski solution~\cite{Horowitz:1997jc} 
instead of obtaining an extremal black hole~\cite{Chen:2021dsw}.\footnote{See also~\cite{Cano:2018hut} that studies the validity of the small black hole analysis.} 
Secondly, there is no known 
AdS$_2 \times S^{d-1}$ solution in supergravity which admits 16 Killing spinors. This could be an issue of simply not having found such a solution but,
in fact, we can sharpen this argument by analyzing possible superconformal groups in the near-horizon region. 
As we explain in Section~\ref{sec:canitbe}, by studying the quantum super Schwarzian 
theory near the horizon, one can rule out the existence of a decoupled near-horizon~AdS$_2$ 
region with 16 or more supercharges describing the heterotic string with all its symmetries. 

\medskip

In order to clarify the situation, it is useful to focus on particular observables
that are amenable to precise calculation. 
In particular, the 
supersymmetric index~\cite{Witten:1981nf} 
is independent of coupling, and can be 
precisely compared in the two descriptions. 
For the $\frac12$-BPS states that we are considering, the relevant index is a helicity supertrace in flat spacetime~(see  e.g.~\cite{Kiritsis:1997gu}). 
This quantity is protected against changes of coupling, for the usual reason of bosonic and fermionic supermultiplets appearing in pairs. 
In the free theory, 
the generating function of this index is the chiral partition function of 24 free bosonic fields 
i.e.~$1/\eta(\tau)^{24}$~\cite{Dabholkar:1989jt}, whose coefficient at level~$nw$ grows as~${\rm exp}\,(4 \pi \sqrt{nw})$ as mentioned above.  
The question we would like to address is: 
what is the saddle-point of the dual gravitational path integral of this index? 

The aim of this paper is to present and analyze this gravitational saddle-point. 
Here, by saddle-point we mean a Euclidean solution of the low-energy supergravity, 
treating~$G_N$ in perturbation theory. 
The main subtlety---and difference from previous treatments---is the inclusion 
of~$(-1)^{\sf F}$ in the gravitational functional integral. We use the 
construction of the saddle-point for the gravitational index discussed in~\cite{Cabo-Bizet:2018ehj,Iliesiu:2021are}. The basic idea is that 
$(-1)^{\sf F}$ can be included in the gravitational functional integral by adding an imaginary part to a 
background bosonic field that couples to an $R$-symmetry. This implies that the solution 
is necessarily complex. The Euclidean solution is non-extremal and has the topology of a cigar times~$S^{d-1}$, with the horizon at the tip of the cigar. 

In asymptotic flat space, 
the relevant background field that implements~$(-1)^{\sf F}$ is the angular velocity~$\Omega$ which obeys~$\beta \Omega = 2 \pi \ii$. 
The solution has non-zero 
rotation along one axis, which defines a north and south pole at the 
horizon. Applying this idea to four-dimensional~$\CN=2$ supergravity coupled to an arbitrary number 
of vector multiplets leads to the \emph{new attractor mechanism}~\cite{Boruch:2023gfn}. 
The metric of these solutions can be written in the Israel-Wilson-Perj\'es (IWP) form~\cite{Perjes:1971gv,Israel:1972vx}, 
which is sourced by harmonic functions with two simple poles at the north and south 
poles, respectively. Near the poles, the values of the scalars of supergravity get fixed in 
terms of the charges of the black hole. The complex nature of the solution is 
reflected in that the holomorphic scalars are fixed at the north pole and vanish at 
the south pole. (The anti-holomorphic scalar has the opposite behavior.)

In this paper we construct the rotating Euclidean black hole as a solution to the 
effective four-dimensional supergravity describing the heterotic string theory on~$T^6$. 
We also write the solution in the variables of the ten-dimensional heterotic theory.
As it turns out, the solution is singular at the two-derivative level. 
In the Einstein frame in four dimensions, the metric is smooth but the string coupling goes to zero and so the string-frame metric is singular. 
This opens the possibility that higher-derivative effects can resolve this 
singularity. 

In order to resolve this singularity, we  study the effect of higher-derivative corrections in the supergravity action on the Euclidean rotating black hole solution. 
We discuss a class of higher-derivative corrections that are summarized in the 
holomorphic prepotential of four-dimensional~$\CN=2$ supergravity, and 
can be calculated using topological string methods in the dual Type II theory. 
In particular, the one-loop term in the prepotential leads to the supersymmetric 
completion of a particular four-derivative term proportional to the square of the Weyl tensor. 
This leads to an extension of the new attractor mechanism to include higher-derivative terms.

In the two-charge system, we find that the singularity is resolved in the higher-derivative theory and, in particular, 
the string coupling at the tip is now non-zero. Note that the resolution involves a play-off between 2-derivative and 4-derivative 
effects. This would lead to an uncontrollable approximation for generic observables. 
However, the calculation for supersymmetric quantities such as the index has a better chance of leading 
to a precise result, along the lines of the non-renormalization 
theorem discussed in~\cite{Butter:2014iwa,Murthy:2013xpa}. 
In particular, if $D$-terms vanish on our supersymmetric configurations, 
the fact 
that the topological string expansion for Type II on $K3 \times T^2$ is one-loop exact implies that 
our result would be rigorous. 
%\footnote{We would like to emphasize that this is a considerable improvement. As opposed to previous work, which could not justify the restriction to F-terms  }. 
%Further, assuming  the action is independent\footnote{} of $\beta$, we evaluate it in the low temperature limit.
Quite remarkably we find that, with these assumptions,  
the gravitational free-energy of the index agrees precisely with that of 
the microscopic index to give the entropy~$4 \pi \sqrt{nw}$ including the numerical coefficient\footnote{To 
evaluate the action we assume it is temperature independent and compute it in the large $\beta$ limit. 
At two derivative level we have proved this earlier~\cite{Boruch:2023gfn}. 
The calculation of the on-shell action (the Gibbons-Hawking free 
energy of the system) at finite $\beta$ including the effect 
of 4-derivative terms is 
complicated and we postpone it to future work.}.

\bigskip

Now we briefly summarize the overall picture that we propose.
First consider the thermal partition function and states in the Lorentzian theory.
Away from extremality the fundamental heterotic string is described by the two-charge black hole solution with a macroscopic horizon. 
As we lower the mass towards extremality there is a transition to a gas of free 
strings, 
without ever reaching the extremal black hole solution~\cite{Chen:2023mbc}. 
Now consider the gravitational path integral for the supersymmetric index as in the present paper. 
The saddle-point is a supersymmetric non-extremal Euclidean geometry valid at all temperatures. 
The saddle-point is really like an instanton rather than describing Lorentzian states. 
At the two-derivative level the solution has a large horizon 
area, but contains curvature singularities 
near the poles of the horizon. These singularities are  
resolved by the inclusion of string-scale higher-derivative corrections in the effective action coming from string theory.  
The index calculated in this manner is independent of coupling, 
and agrees with the index of a gas of strings at weak-coupling.\footnote{There may even exist a Horowitz-Polchinski-type supersymmetric solution at intermediate coupling. The 
details of this would be interesting and may involve a perhaps suitably complexified sigma-model. We comment further on this possibility in Section~\ref{sec:impsusy}.} 
In this respect, the index behaves in a manner similar to the supersymmetric indices corresponding to big black holes\footnote{Big black holes refer to black holes with with 
horizon curvatures parameterically smaller than the string and Planck scale.}~\cite{Cabo-Bizet:2018ehj,Iliesiu:2021are,Boruch:2023gfn}. They do not depend on coupling, and have continuously connected dual pictures in the gravitational and microscopic regimes, rather than a sharp transition. 

\bigskip

The plan of paper is as follows. 
In Section~\ref{sec:SBHthermo} we review the family of two-charge black hole 
solutions in heterotic string theory, their thermodynamics, and the transition to the 
gas of free strings via the Horowitz-Polchinski solution. 
In Section~\ref{sec:2dernonextsusy} we discuss the gravitational saddle-point 
for the index, i.e.~the Euclidean rotating black hole, in the heterotic string variables.
In Section~\ref{sec:4dsugra} we switch to the four-dimensional effective supergravity
description at the two-derivative level. We review the new attractor mechanism and the embedding of the Euclidean rotating black hole in this description. 
In Section~\ref{sec:higherdersol} we include a four-derivative term in the 4d supergravity coming from the string 
compactification. We show how it desingularizes the Euclidean rotating black hole, and analyze its effect on the entropy. 
In Section~\ref{sec:canitbe} we analyze the superconformal groups that can possible arise in the near-horizon geometry 
and the quantum effects of the respective super-Schwarzians.  We show that AdS$_2$ cannot arise at the quantum level in the two-charge system. 
In Section~\ref{sec:discussion} we discuss further possible directions. 
In the appendix we discuss how the solution 
generating technique~\cite{Narain:1985jj,Narain:1986am} relates the rotating charged and uncharged solutions. 

\medskip

\paragraph{Note added} While this paper was in preparation we received~\cite{Chowdhury:2024ngg}, which 
contains overlap on the two-derivative solutions that calculate the index.

\section{The two-charge black hole and its transition to free strings}\label{sec:SBHthermo}

In this section we review the properties of the two-charge black hole in the heterotic string theory on $T^6$. 
In particular, we discuss some peculiar features of the naive extremal limit of this solution. 
Then we review the construction of the charged Horowitz-Polchinski solution and how it modifies the approach to extremality.

\subsection{Review of the two-charge black hole} 

We begin by reviewing the two-charge black hole, constructed by Sen in~\cite{Sen:1994eb}, as a solution to the four dimensional supergravity theory 
arising from a toroidal compactification of heterotic string theory. 

\subsubsection{Heterotic string on $T^6$}

The four dimensional theory has $\mathcal{N}=4$ supersymmetry and contains the supergravity multiplet, and~22 $\mathcal{N}=4$ vector multiplets.
The supergravity multiplet has six graviphotons, resulting in a total gauge group $\U(1)^{28}$. 
We collectively denote the potentials by~$A^a$ with $a=1,\ldots, 28$. 
The moduli space consists of the dilaton $\Phi$ and 132 scalar fields organized in a symmetric $28\times 28$ dimensional matrix~$M$ that satisfies $M L M^T = L$. 
$L$ is a diagonal matrix $L_{aa}=-1$ for $a=1,\ldots,22$ and $L_{aa}=+1$ for $a=23,\ldots,28$. 
The scalars $M$ encode deformations of the torus, of the background $B$-field, and of the Wilson lines. 
Besides these fields, we also have a four dimensional $B$-field $B_{\mu\nu}$, which is dual to a scalar.

The action, in Lorentzian signature, is given by 
\bea\label{action}
I &\=& \frac{1}{16\pi}\int \d^4 x\, \sqrt{-g} \, e^{-\Phi}\, \Big[ R + (\partial \Phi)^2 + \frac{1}{8} {\rm Tr} \left[ \partial_\mu M \, L \, \partial^\mu M \, L \right]\nn
&& ~~~- \frac{1}{12} H_{\mu\nu\rho} H^{\mu\nu\rho} - F^a_{\mu\nu}\, (L ML)_{ab} \, F^b{}^{\mu\nu} \Big]+({\rm fermions })+({\rm bdy~terms})\,,
\ea
where we denote the $\U(1)^{28}$ field strength by~$F^a_{\mu\nu} = \partial_\mu A_\nu^a - \partial_\nu A_\mu^a $ 
and the $B$-field field strength by $H_{\mu\nu \rho} = \partial_\mu B_{\nu\rho} + 2 A^a_\mu L_{ab} F^b_{\nu \rho} + ({\rm cyclic})$. 
The action is invariant under an $O(6,22,\mathbb{R})$ transformation which acts on $a,b$ indices only, 
so it does not act on the dilaton, metric or $B$ field. 
Consistency with quantization of electromagnetic charge restricts it to $O(6,22;\mathbb{Z})$.

The solutions we discuss in the following have boundary conditions at infinity such that locally
\beq
M_{ab}|_{\infty} \= \delta_{ab}\,, \qquad \Phi |_{\infty} \= 0\,, \qquad g_{\mu\nu} |_{\infty} \= \eta_{\mu\nu} \,.
\eeq
For the gauge fields, both the electromagnetic ones and the $B$-field, we can choose to fix their holonomies or their charges at infinity.

\subsubsection{Sen's black hole solution}

The procedure to find the electrically charged black holes described in \cite{Sen:1994eb} is the following. 
First, one realizes that for metrics with a $\U(1)$ isometry (corresponding to Euclidean time), the symmetry of the action gets enhanced to $O(7,23)$. 
One can think about this in terms of a $T^7= T^6 \times S^1 $ compactification of the heterotic string, but \cite{Sen:1994eb} verifies it by explicit construction of the $O(7,23)$ action. 
We discuss such transformations more generally in Appendix~\ref{sec:solngene}, not necessarily restricting our attention to black hole solutions. 

Let us present the final results. The metric in the Einstein frame $g_{\mu\nu}^E=e^{-\Phi} g_{\mu\nu}$  and Lorentzian signature is given by 
\bea\label{Sensoln}
e^{-\Phi} \, \d s^2 &\=&\Delta^{1/2} \Big[ - \frac{r^2 +a^2 \cos^2 \theta - 2 m r}{\Delta} \d t^2 + \frac{\d r^2}{r^2 + a^2 - 2 m r} + \d \theta^2  \nn
&& \qquad + \frac{\sin^2 \theta}{\Delta} [ \Delta + a^2 \sin^2\theta(r^2 + a^2 \cos^2 \theta+ 2 m r \cosh \upalpha \cosh \upbeta)] \,\d \phi^2\nn
&&  \qquad - \frac{2 m r a \sin^2 \theta (\cosh \upalpha + \cosh \upbeta)}{\Delta}\, \d t \d \phi \Big] \,,
\ea
where we define the function
\beq
\Delta \= (r^2 + a^2 \cos^2 \theta)^2 + 2 m r ( r^2 + a^2 \cos^2 \theta) (\cosh\upalpha \cosh\upbeta -1) + m^2 r^2 (\cosh \upalpha - \cosh \upbeta)^2 \,.
\eeq
The dilaton is given by 
\beq
\Phi \= \frac{1}{2} \log \frac{(r^2 + a^2 \cos^2 \theta)^2}{\Delta}.
\eeq
We can decompose the vector fields into a 22-dimensional vector of one-forms $\vec{A}_L$, made up of the first 22 components of $A$, 
and a 6-dimensional vector of one-forms $\vec{A}_R$ made of the last 6 components of $A$. 
These gauge potentials and the two-form are given by
\bea
\vec{A}_L &\=& - \frac{\vec{n}}{\sqrt{2}} \frac{m r \sinh \upalpha ((r^2 + a^2 \cos^2 \theta)\cosh\upbeta + m r (\cosh\upalpha-\cosh\upbeta))}{\Delta} \d t \nn
&&+ \frac{\vec{n}}{\sqrt{2}} \frac{ m r a \sinh \upalpha \sin^2 \theta  (r^2 + a^2 \cos^2 \theta + m r \cosh\upbeta (\cosh\upalpha-\cosh\upbeta))}{\Delta} \d \phi\,,\\
\vec{A}_R &\=& - \frac{\vec{p}}{\sqrt{2}} \frac{m r \sinh \upbeta ((r^2 + a^2 \cos^2 \theta)\cosh\upalpha + m r (\cosh\upbeta-\cosh\upalpha))}{\Delta} \d t \nn
&&+ \frac{\vec{p}}{\sqrt{2}} \frac{ m r a \sinh \upbeta \sin^2 \theta  (r^2 + a^2 \cos^2 \theta + m r \cosh\upalpha (\cosh\upbeta-\cosh\upalpha))}{\Delta} \d \phi\,,\\
B&\=&\frac{m r a \sin^2 \theta (\cosh\upalpha- \cosh\upbeta)}{\Delta}( r^2 + a^2 \cos^2 \theta + m r (\cosh\upalpha \cosh\upbeta -1)) \,\d t  \d \phi\,.
\ea
The scalar moduli are given by
\beq
M \= I_{28} + \begin{pmatrix} 
{\sf P} n n^T  & {\sf Q} n p^T \\
{\sf Q} p n^T & {\sf P} p p^T 
\end{pmatrix}\,,
\eeq
where 
\bea
{\sf P} &\=& \frac{2 m^2 r^2 \sinh^2 \upalpha \sinh^2 \upbeta}{\Delta},\\
{\sf Q} &=& -\frac{2 m r \sinh \upalpha \sinh\upbeta}{\Delta}(r^2 + a^2 \cos^2 \theta + m r (\cosh\upalpha \cosh\upbeta -1))\,.
\ea
Here $\vec{n}$ and $\vec{p}$ are arbitrary 22 and 6 dimensional unit vectors. 
It will be important to know the charges and chemical potentials as functions of the parameters appearing in the solution, 
which are~$(m,a,\upalpha,\upbeta, \vec{n},\vec{p})$. 
Define~$\vec{Q}_L $ to be a 22-dimensional vector made out of the first 22 components of $Q^a$. 
Similarly define~$\vec{Q}_R$ to be the last 6 components.  
To compute the charges we can evaluate the large~$r$ limit of the gauge potentials and read them off from~$A \sim Q/r\, \d t$. 
Using that $\Delta \sim r^4 $ as~$r\to\infty$ we obtain
\beq\label{QLQR}
\vec{Q}_L \= \frac{m}{\sqrt{2}} \sinh \upalpha \cosh \upbeta \, \vec{n} \,, \qquad \vec{Q}_R= \frac{m}{\sqrt{2}} \sinh \upbeta \cosh\upalpha \, \vec{p}\,.
\eeq
We will also use $Q_L, Q_R$ to denote the modulus of the respective charge vectors. 
Since the black holes are electrically charged and rotating there will also be a non-trivial magnetic dipole moment, 
but it will not be important for the discussion here. 

Let us now move on to the other charges encoded in the metric. The mass and angular momentum of the 
black hole can be read off from the large~$r$ behavior of~$g_{tt}$ and~$g_{t\phi}$ and are given by
\beq\label{MJ}
M \= \frac{m(1+ \cosh \upalpha \cosh \upbeta)}{2} \,, \qquad J \= \frac{m a (\cosh \upalpha + \cosh \upbeta)}{2}.
\eeq
The metric has an outer horizon~$r_+$ at the locations where~$g_{rr}^{-1}$ vanishes
\beq\label{twohor}
r^2 - 2 m r + a^2 = 0 \quad \Rightarrow \quad r_\pm \= m \pm  \sqrt{m^2 - a^2}\,.
\eeq
The temperature and angular velocities are
\bea
\beta &\=& \frac{2\pi m (\cosh \upalpha + \cosh \upbeta)(m+\sqrt{m^2-a^2})}{\sqrt{m^2-a^2}}\,,\\
\Omega &\=& \frac{a}{m(\cosh \upalpha + \cosh \upbeta)(m+\sqrt{m^2-a^2})}\,.
\ea
We see that for the Lorentzian solution to make sense we need $m\geq a$ and the limit $m\to a$ is an extremal limit 
(in the sense that inner and outer horizons have the same area) although the temperature \textit{does not} vanish. 
We will expand on this later. When $a\to0$ and $\upalpha, \upbeta\to 0$ then the solution 
becomes the Schwarszchild black hole. 
We can check this by looking at the potentials, for example $\beta \to 8 \pi m$. 

Having found the location of the horizon and the angular velocity, we can evaluate the electric chemical potentials. 
These quantities are given by $\mu = {\sf i}_V A |_{r\to\infty} -{\sf i}_V A |_{\rm hor.}$ where $V= \partial_t + \Omega \partial_\phi$. 
This is a gauge invariant statement, but to be more rigorous we should 
apply a large gauge transformation such that ${\sf i}_V A |_{\rm hor.} = 0$.  
The electric potentials are
\beq
\vec{\mu}_L \= \frac{\vec{n}}{\sqrt{2}} \frac{\sinh \upalpha}{\cosh\upalpha + \cosh \upbeta}\,, \qquad \vec{\mu}_R \= \frac{\vec{p}}{\sqrt{2}} \frac{\sinh \upbeta}{\cosh\upalpha + \cosh \upbeta}\,.
\eeq
Finally we can compute the area of the horizon, 
with respect to the Einstein frame metric,
\begin{equation}\label{eq:area}
    A \= 4 \pi m (\cosh \upalpha + \cosh \upbeta) (m + \sqrt{m^2 -a^2}).
\end{equation}
The on-shell action of the black hole is, in the grand canonical ensemble where we set Dirichlet boundary conditions for  all fields, given by 
\beq
\log Z \= - I_{\rm on-shell}\= \frac{A}{4} - \beta M + \beta \vec{\mu}_L \cdot \vec{Q}_L + \beta \vec{\mu}_R \cdot \vec{Q}_R + \beta \Omega J \,.
\eeq

\medskip

We can compare the supersymmetric limit with the extremal limit. We already mentioned the extremal limit is $m=a$. The solution is supersymmetric when
\beq\label{SUSY}
{\rm SUSY}: \qquad M^2 \= \frac{\vec{Q}_R^2}{2}.
\eeq
Combining with (\ref{MJ}) and (\ref{QLQR}) we see that it implies 
\beq\label{SUSYeqn}
m^2(1+\cosh\upalpha \cosh \upbeta)^2 \= m^2\cosh^2 \upalpha \sinh^2 \upbeta \,.
\eeq
If all parameters are finite and real we cannot satisfy (\ref{SUSYeqn}) as we need either (i) $m=0$, implying $M=0$ and $\vec{Q}_R=\vec{0}$ and, moreover, $A=0$---implying a singularity at the horizon; 
or (ii) we have $m\neq 0$ but the solution is complex, for example we can see that if we solve 
for $\cosh \upalpha$ as a function of $\cosh \upbeta$ then $\cosh \upalpha<1$ for 
all $\cosh\upbeta \in (+1,\infty)$. 
As explained by Sen the supersymmetric solutions have $m\to 0$ and either $\upbeta$ going to infinity or $\upalpha,\upbeta$ 
going to infinity together, but it is not always extremal in the sense of being zero temperature.

\subsubsection{Thermodynamics without $(-1)^{\sf F}$ insertion}\label{sec:naivethermo}

We want to describe the termodynamics of non-rotating electric black holes in this setup, without any insertion 
of~$(-1)^{\sf F}$ (which we will discuss in later sections). We then 
set~$a=0$ and the expressions in the previous section simplify. 
The mass and entropy of the black hole are given by
\begin{equation}\label{MS}
\begin{aligned}
  &  M = \frac{m(1+ \cosh \upalpha \cosh \upbeta)}{2}, \quad S = 2\pi m^2 (\cosh \upalpha + \cosh \upbeta), 
\end{aligned}
\end{equation}
and the temperature 
\beq\label{betanoQ}
\beta = 4\pi m (\cosh \upalpha + \cosh \upbeta).
\eeq
The expressions for the charges and electric potentials are unmodified. 

There are several distinct limits one can consider. 
In this section we assume the validity of~(\ref{MS}) 
and~(\ref{betanoQ}), and discuss the thermodynamics implied by them. 
However, we should keep in mind that these relations have~$\alpha'$ corrections which can become important when the black hole 
becomes small and, in fact, as we will discuss in 
Section~\ref{sec:HP}, the correct description in certain regime is in fact given by horizonless geometries. 

\paragraph{Low temperature limit} The simplest limit one can consider is the low temperature limit, i.e. $\beta \to\infty$, while holding the charges fixed. In this limit we have
\bea
&&\sinh \upalpha \;\sim\; \sqrt{2} \frac{8 \pi Q_L}{\beta} \;\sim\; 0 \,, \qquad \cosh \upalpha \;\sim\; 1 \,,\\
&&\sinh \upbeta \;\sim\; \sqrt{2} \frac{8 \pi Q_R}{\beta} \;\sim\; 0 \,, \qquad \cosh \upbeta \;\sim\; 1 \,.
\ea
For this reason, if we compute the mass and entropy as functions of temperature we get
\beq
M \;\sim\; \frac{\beta}{8\pi}\,, \qquad 
S \;\sim\; \frac{\beta^2}{16 \pi}.
\eeq
We simply recover the behavior of a regular Schwarszchild black hole, the bigger the black hole the bigger the mass and the colder it gets. 
Since the black hole becomes large and weakly curved, the black hole description is trustworthy even though the ensemble is thermodynamically unstable. 

\paragraph{The naive extremal limit} Since we are considering here the special case~$a=0$, what we mean by the extremal limit is~$m\rightarrow 0$. 
However, to keep the charges (\ref{QLQR}) fixed, the limit is not so straightforward. The only solution to this problem is to take either $|\upalpha|\to \infty$ or~$|\upbeta| \to \infty$. 
For simplicity, here we will focus on the cases where~$\upalpha,\upbeta>0$. The generalization of the discussion to the cases where one or both of them are negative is straightforward.

\begin{figure}[t!]
\centering
\begin{tikzpicture}
\pgftext{\includegraphics[scale=0.3]{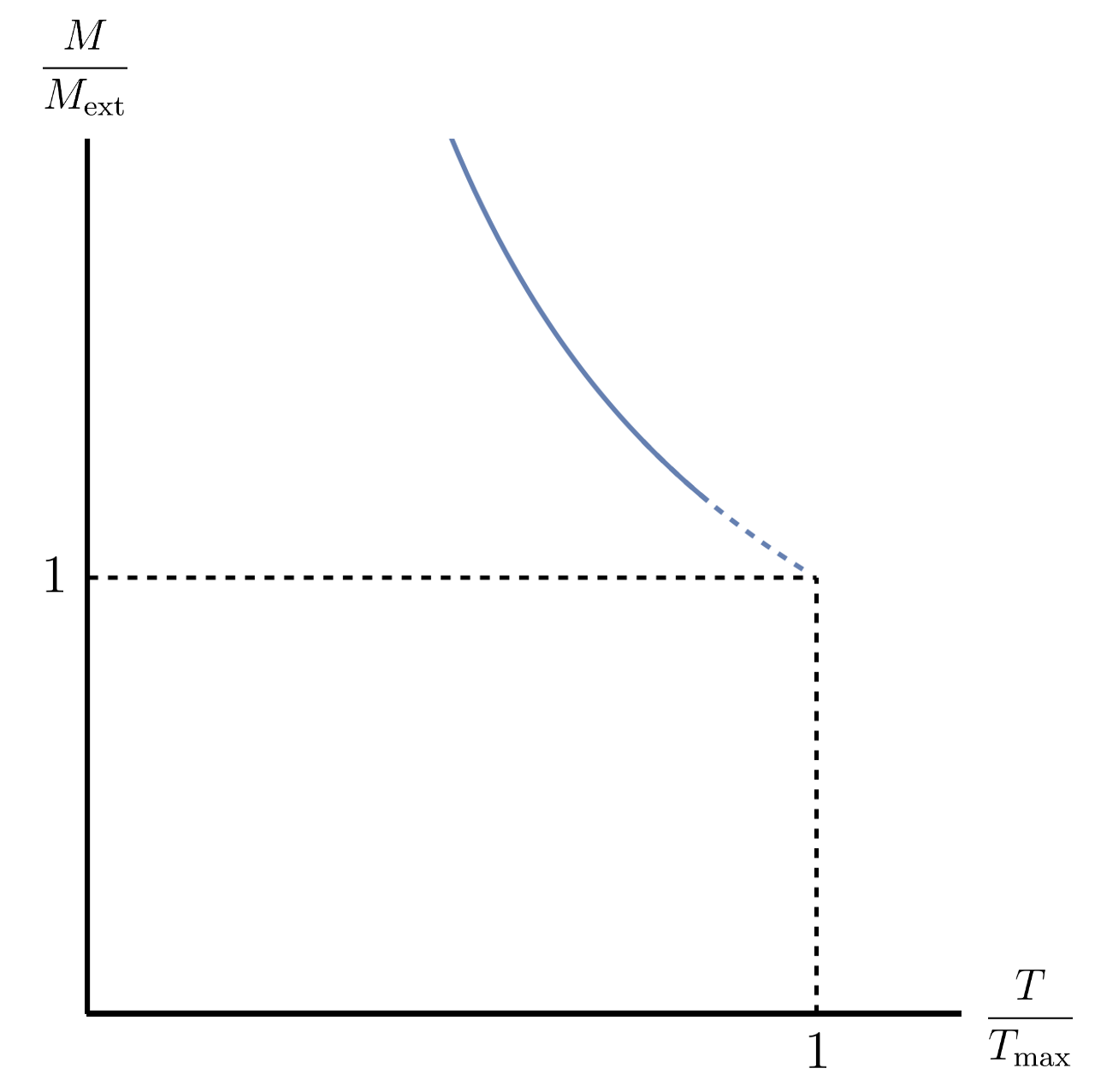}} at (0,0);
\draw (0,-3.5) node {\footnotesize $(a)$};
\end{tikzpicture}
\quad\quad\quad\quad
\begin{tikzpicture}
\pgftext{\includegraphics[scale=0.3]{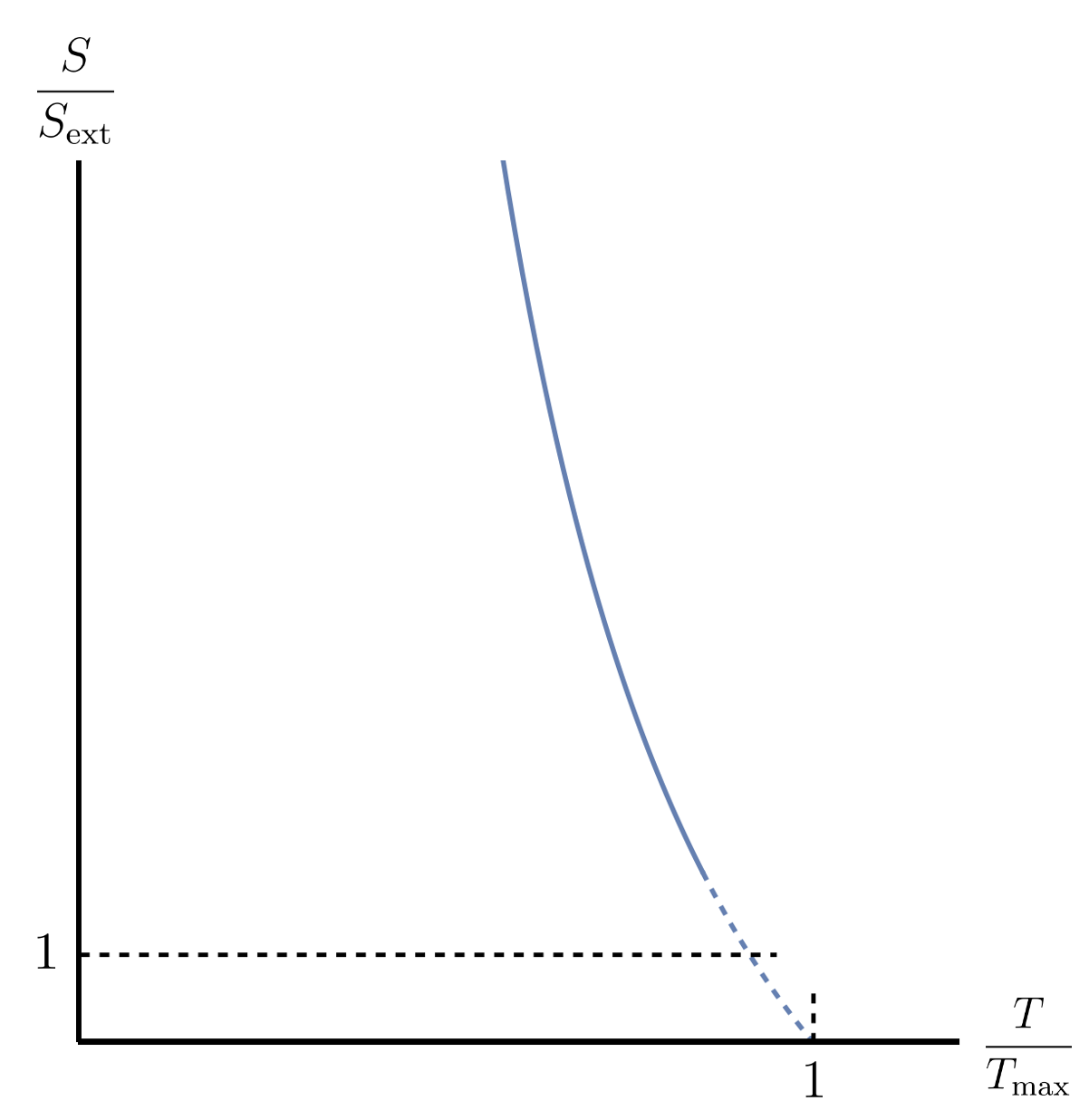}} at (0,0);
\draw (0,-3.5) node {\footnotesize $(b)$};
\end{tikzpicture}
\caption{\footnotesize (a) An illustration of the black hole mass (divided by its extremal value $M_{\textrm{ext}}$) as a function of temperature (divided by $T_{\textrm{max}} =1/(4\pi \sqrt{2}\sqrt{| Q_R^2 - Q_L^2|})$). The mass decreases until the temperature reaches maximum. The dashed part of the blue line means that the black hole is of string scale and the curve is just an extrapolation.   (b) An illustration of the black hole entropy (divided by the extremal entropy $S_{\textrm{ext}}$ from microstate counting) as a function of temperature. The naive extrapolation of the entropy simply vanishes when we reach the maximum temperature. }
\label{fig:QL=0}
\end{figure}

We will consider first the case of $\upbeta \to \infty$, which as it turns out corresponds to 
the case of $Q_R > Q_L$. To keep the charges fixed, we take $\upbeta \to\infty$ and $m \to 0$ while keeping
\beq
m \cosh \upbeta \to m_0
\eeq
finite. The first surprise, under the assumption that we can trust the black hole solution, 
is that this extremal limit does not imply vanishing temperature. Instead 
\beq\label{betaext}
\beta_{\rm ext} \= 4 \pi m_0\,,\qquad Q_L \= \frac{m_0}{\sqrt{2}} \sinh \upalpha\,, \qquad Q_R \= \frac{m_0}{\sqrt{2}} \cosh \upalpha.
\eeq
From the last two equations of (\ref{betaext}) we see that $\tanh \upalpha = Q_L/Q_R$ and $m_0 =\sqrt{2}\sqrt{Q_R^2 - Q_L^2}$. Therefore the extremal limit has a finite, charge dependent, temperature
\begin{equation}
   T_{\textrm{max}} \= \frac{1}{4\sqrt{2}\pi \sqrt{Q_R^2 - Q_L^2}}.
\end{equation}
This solution is precisely Sen's small black hole since the area of the horizon now vanishes
\beq
A \;\sim\; m^2 \cosh \upbeta \sim m_0 m \to 0.
\eeq
We saw earlier that at low temperature the black hole is very large since it behaves as the Schwarszchild solution. 
As the temperature increases the mass decreases until it reaches extremality at $\beta_{\rm ext}$. 
This is the situation if $Q_R > Q_L$. At this point the entropy vanishes since the area of the horizon vanishes. 
If we further extrapolate the expressions to temperature higher than~$T_{\textrm{max}}$, 
the mass parameter becomes negative $m<0$, leaving a naked singularity. 
Moreover, the parameter~$\upbeta$ goes to the complex plane. 
At the temperature~$T_{\textrm{max}}$ the value of the mass is given 
by~$M_{\textrm{ext}}^2 = Q_R^2 /2$ which is precisely the BPS mass 
which is the lowest possible from the dual quantum mechanical point of view.  
In Figure~\ref{fig:QL=0}, we show how the mass and entropy behave as functions of temperature. 
We should however stress that we should not trust the curves when the temperature becomes too close to $T_{\rm max}$ and the black hole becomes string scale. 
We will discuss the correct behavior in Section~\ref{sec:HP}.

Above we considered the case of $\upbeta\rightarrow \infty$ and therefore $Q_R > Q_L$. The discussion of the case $Q_{L}> Q_R$ essentially mirrors the previous case, but the interpretation is slightly different. In this case as the black hole temperature rises, the solution eventually has $\upalpha \to \infty$ instead. The expressions above are still true, but now the lowest mass black hole has $M= Q_L /\sqrt{2}$ which is not supersymmetric. Since $Q_L > Q_R$ there is now a range of masses, namely $Q_L^2/2 \geq M^2 \geq Q_R^2 /2$ which is allowed by unitarity in the microscopic theory, but where black holes do not exist. Therefore depending on the sign of $Q_R-Q_L$ the ground state might be BPS or not. 

Finally, if $Q_L=Q_R$ then $\upalpha = \upbeta$. We will not go into the details of this special case, but just  note that we need to take both of them to infinity, 
and the extrapolation of~(\ref{betanoQ}) suggests that the extremal limit corresponds to infinite temperature $\beta_{\rm ext} = 0$. 

We note that the black hole solution in the two derivative theory (\ref{action}) is symmetric with respect to turning on the left or right moving charges. However, from the microscopic theory point of view, we wouldn't expect such symmetry. 
For one, the solution should be supersymmetric only 
when~$M=Q_R/\sqrt{2}$ but not when~$M=Q_L/\sqrt{2}$. As it turns out, this symmetry is broken once one considers the leading $\alpha'$ corrections to the black hole~\cite{Massai:2023cis}.

\subsection{Transition to free strings via the charged Horowitz-Polchinski solution}\label{sec:HP}

As we saw in the previous section, the extrapolation of the gravity solution~(\ref{Sensoln}) to the extremal limit has some peculiar features. The black hole solution becomes singular as the horizon shrinks to zero size. Of course, once the size of the horizon becomes comparable to string scale, the higher curvature corrections would become important and there is no reason to trust the solution~(\ref{Sensoln}). In this section, we will review the result in~\cite{Chen:2021dsw}, which states that the approach to extremality of the two-charge system is in fact described by a charged analogue of the Horowitz-Polchinski solution~\cite{Horowitz:1997jc}. 

Here we will only sketch the idea behind the construction of the solution as details can be found in \cite{Chen:2021dsw}. The original uncharged Horowitz-Polchinski solution can be constructed explicitly by considering the effective action of the thermal winding modes coupled to gravity and dilaton. 
The uncharged solution is trustworthy when the temperature is below but very close to the Hagedorn temperature, $(\tilde{\beta} - \beta_H)/ \beta_H \ll 1$,\footnote{We will use tilded quantities to denote the quantities in the uncharged solution, which are distinct from those of the charged solution but related via the solution generating procedure, see Appendix. \ref{sec:solngene} for a more detailed discussion.} where we can focus on only the winding modes with winding number $\pm 1$ and the fluctuation of the $g_{tt}$ component of the metric. 
The solution carries a classical entropy, which agrees with the Hagedorn density of states for a gas of strings close to the Hagedorn temperature. This suggests that the physical interpretation of the solution is describing a gas of highly excited strings, weakly interacting via gravity (see also \cite{Damour:1999aw}). It is further suggested in \cite{Chen:2021dsw} that in Heterotic string theory, this classical solution could potentially be smoothly connected to the Schwarzschild black hole solution as we lower the temperature.\footnote{While in the Type II theories, a worldsheet index argument suggests that the two solutions cannot be smoothly interpolated.} 

An important property of the solution which differs from the Schwarzschild black hole is that as we increase the temperature, the physical size $ \ell_{\textrm{size}} $ of the solution expands as $1/\sqrt{\tilde{\beta} - \beta_H}$. Physically, this means the classical solution is describing a more and more dilute string gas. When the temperature gets so close to the Hagedorn temperature such that the size of the classical solution approaches the size of a free string gas with the same mass, the classical solution breaks down and the correct description is simply the quantum description of free string gas. In four dimensions, the solution is only valid when
\begin{equation}\label{valid}
    \tilde{g}_s^{\frac{4}{3}} \lesssim  (\tilde{\beta} - \beta_H)/ \beta_H \ll 1.
\end{equation}

Given the uncharged solution, \cite{Chen:2021dsw} performed a solution generating transformation to find the charged solution. The explicit form of the solution can be found in \cite{Chen:2021dsw}, while here we will only mention a few main properties. It turns out that the $\tilde{\beta} \rightarrow \beta_H$ limit of the ``seed" solution is mapped to the extremal limits $M \sim Q_R/\sqrt{2}$ or $M\sim Q_L/\sqrt{2}$ of the generated solutions. Differing from the naive extrapolation of the black hole results in Section~\ref{sec:naivethermo}, where the temperature rises and reaches a maximum, the temperature of the generated Horowitz-Polchinski solution goes to zero, i.e. $\beta\rightarrow \infty$ in the extremal limit. Furthermore, the solution correctly reproduces the entropy as expected from counting microstates of long strings carrying momentum and winding in the extra directions
\begin{equation}
\begin{aligned}
   &  S \;\sim\;  \sqrt{2} \pi \ell_s \sqrt{Q_R^2 -Q_L^2} \,, \qquad & M \;\sim\; \frac{Q_R}{\sqrt{2}} \,, \\
   & S \;\sim\;  \pi \ell_s \sqrt{Q_L^2 -Q_R^2} \,, \qquad & M \;\sim\; \frac{Q_L}{\sqrt{2}} \,.
\end{aligned}
\end{equation}

The thermodynamics of the charged solution can be determined purely from the thermodynamics of the seed solution plus the form of the solution generating transformation at spatial infinity. We will discuss the derivation of such relations in Appendix~\ref{sec:solngene}, extending the analysis in \cite{Chen:2021dsw} to the cases where the seed solution has rotation. Specifying 
(\ref{geneHet1})--(\ref{geneHet5}) to the case without rotation, and expanding the temperature of the seed solution around $\beta_{H} = 2\pi R_{H} = 2\pi (1+1/\sqrt{2}) \ell_s$, we get %\JT{G-Newton is one?}
\begin{equation}\label{genecharges}
\begin{aligned}
	M & \= \frac{\tilde{S}'}{2\pi R_H} \cosh \upalpha \cosh \upbeta \,, \\
	S & \= \frac{\tilde{S}'}{2 } \left[\left(1 - \frac{\alpha'}{2R_H^2}\right) \cosh \upalpha + \left(1 + \frac{\alpha'}{2R_H^2}\right) \cosh \upbeta \right]\,, \\
	\beta &  \= \frac{\beta_H }{2} \left[\left(1 + \frac{\alpha'}{2R_H^2}\right) \cosh \upalpha +  \left(1 - \frac{\alpha'}{2R_H^2}\right) \cosh \upbeta \right] \,, \\
	Q_L &\= \frac{\tilde{S}'}{\sqrt{2}\pi R_H} \cosh \upbeta \sinh \upalpha \,,\\
	Q_R & \= \frac{\tilde{S}'}{\sqrt{2} \pi R_H} \cosh \upalpha \sinh \upbeta \,,
\end{aligned}	
\end{equation}
where $\tilde{S}'$ is the entropy of the seed solution after stripping off the dependence on the string coupling, and is given by
\begin{equation}\label{Stp}
    \tilde{S}' \;\approx \; c_0 \sqrt{ \ell_s^3 (\tilde{\beta} - \beta_H )}.
\end{equation}
Here $c_0 \approx 2.41$ is an order one number coming from the properties of the seed solution.

In summary, (\ref{genecharges}) and (\ref{Stp}) describe the thermodynamic quantities of the charged Horowitz-Polchinski solution, parameterized by three parameters $\{ \tilde{\beta},  \upalpha,\upbeta\}$. They are only valid when we satisfy (\ref{valid}). Note that the lower end of the inequliaty in (\ref{valid}) is expressed in terms of the string coupling constant in the seed solution, rather than the generated solution. 

With (\ref{genecharges}), we can reexamine how the naive approach to extremality in section  \ref{sec:naivethermo} gets modified. As we mentioned, now the extremal limit corresponds to $\tilde{\beta} \rightarrow \beta_H$, and to fix the charges, we need to combine it with the limits $\upalpha\rightarrow \infty$ or $\upbeta \rightarrow \infty$. For the case of $\upalpha\rightarrow \infty$, we have $M\rightarrow Q_L/\sqrt{2}$, and for $\upbeta \rightarrow \infty$, we have $M\rightarrow Q_R/\sqrt{2}$.  From the expression of $\beta$ in (\ref{genecharges}), we clearly see that both would correspond to the limit where $\beta\rightarrow \infty$, i.e. the temperature goes to zero in the extremal limit. 

We can consider some special cases for which we can write down simpler formulas for how the energy and entropy depends on the temperature. Consider the case $Q_L = 0$ ($\upalpha = 0,\, \upbeta \rightarrow \infty$), at low temperature we have
\begin{equation}
    \frac{M-M_{\textrm{ext}}}{M_{\textrm{ext}} } \;\sim\;  \frac{S-S_{\textrm{ext}}}{ S_{\textrm{ext}}} \;\sim\;  2 e^{-2\upbeta} \;\sim\; (\pi\ell_s T)^2,
\end{equation}
where $M_{\textrm{ext}} = Q_R/\sqrt{2}$ and $S_{\textrm{ext}} = \sqrt{2} \pi \ell_s Q_R$. Translating (\ref{valid}) into the parameters of the generated solution, we find that the solution is trustworthy when
\begin{equation}\label{validity}
    \left(  \frac{\ell_s}{Q_R} \right)^{\frac{1}{2}} \ll  Q_R T \ll 1 \,.
\end{equation}
On the other hand, for the case of $Q_R = 0$ ($\upbeta = 0, \,\upalpha \rightarrow \infty$), at low temperature we have
\begin{equation}
 \frac{M-M_{\textrm{ext}}}{M_{\textrm{ext}} } \;\sim\; \frac{S - S_{\rm ext}}{ S_{\rm ext}} \;\sim\;2 e^{-2\upalpha} \sim 2 (\pi \ell_s T)^2 \,,
\end{equation}
where $M_{\textrm{ext}} = Q_L/\sqrt{2}$ and $S_{\textrm{ext}} = \pi \ell_s Q_L$. The region of validity of the solution is similar to (\ref{validity}), but with $Q_R$ replaced by $Q_L$. Note that we have $\delta S \propto \delta M$, which is expected from the microscopic entropy formula
\begin{equation}
    S \= 2\pi \ell_s \left[\sqrt{ M^2 - \frac{Q_L^2}{2}} + \frac{1}{\sqrt{2}}  \sqrt{ M^2 - \frac{Q_R^2}{2}}\right]
\end{equation}
when we expand around extremality.
\begin{figure}[t!]
\centering
\begin{tikzpicture}
\pgftext{\includegraphics[scale=0.3]{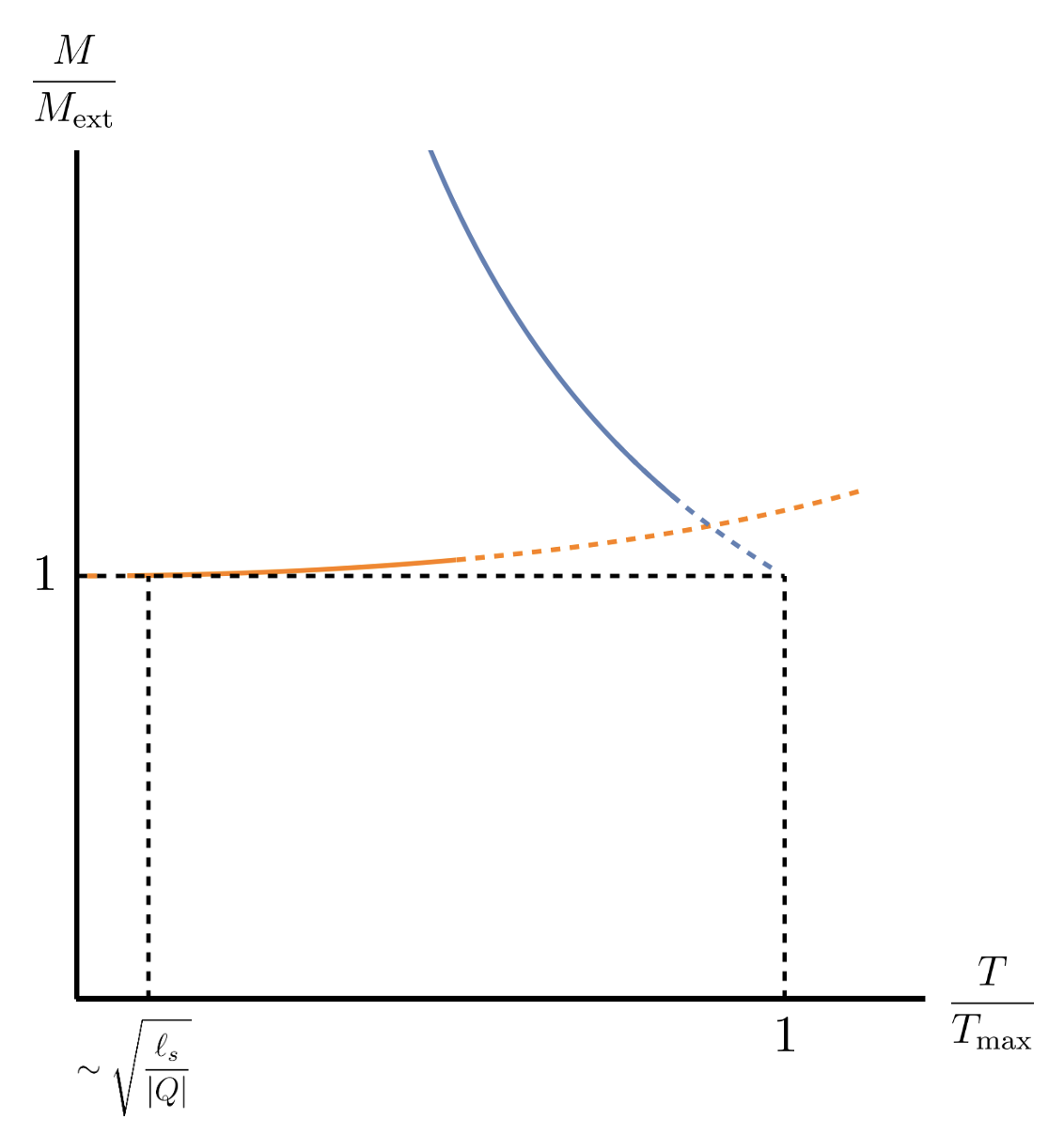}} at (0,0);
\draw (0,-3.5) node {\footnotesize $(a)$};
\end{tikzpicture}
\quad\quad\quad\quad
\begin{tikzpicture}
\pgftext{\includegraphics[scale=0.3]{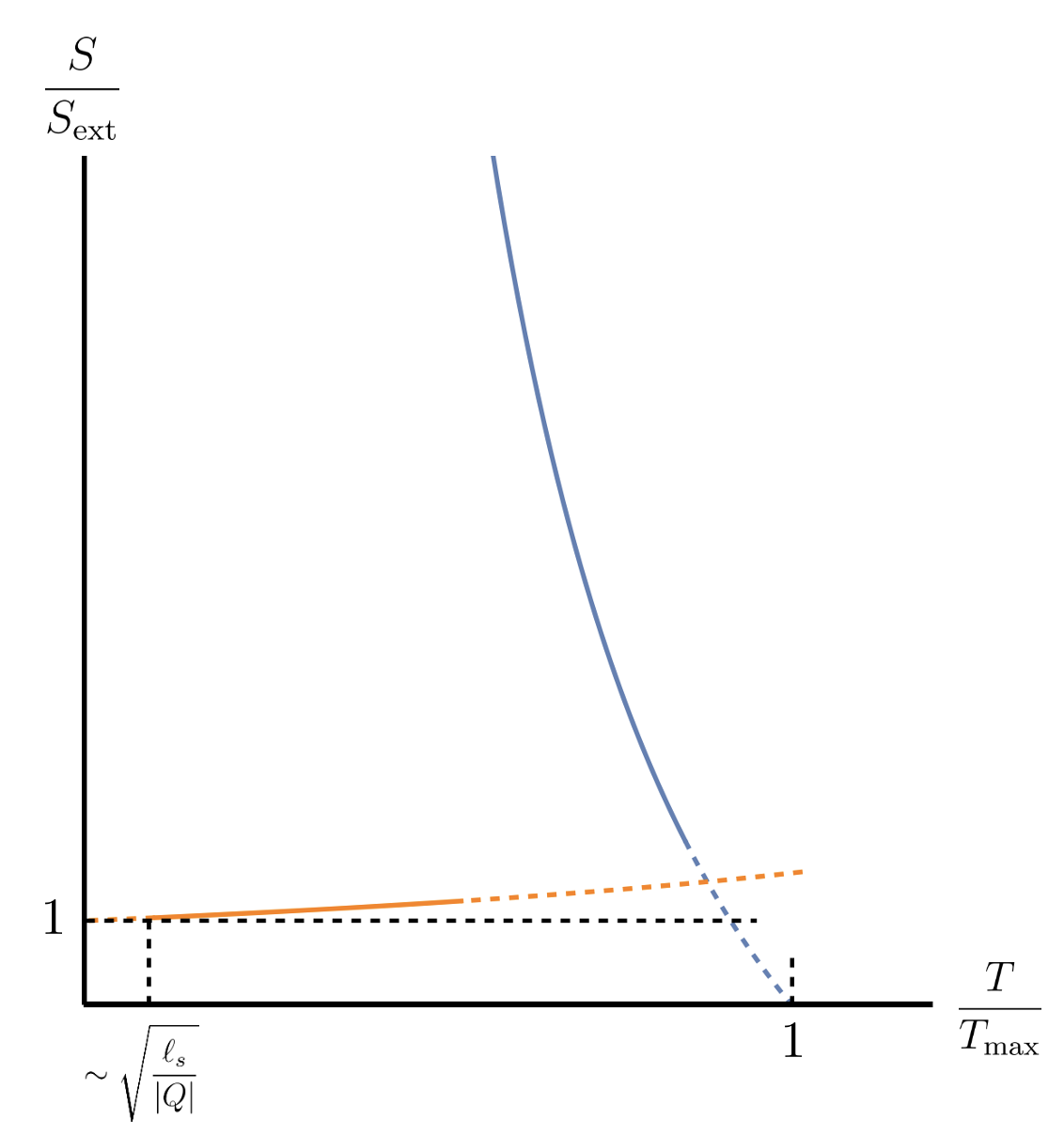}} at (0,0);
\draw (0,-3.5) node {\footnotesize $(b)$};
\end{tikzpicture}
\caption{\footnotesize (a)  The blue curves are from Figure~\ref{fig:QL=0} for comparison. The orange curve is the mass of the charged Horowitz-Polchinski solution as a function of the temperature. It is trustworthy when $\sqrt{\ell_s / |Q|} \ll T/T_{\textrm{max}} \ll 1$, where $|Q|$ can be taken to be the greater of $|Q_L|,|Q_R|$. The dashed lines are extrapolations. We see that the temperature goes to zero in the extremal limit. (b) The orange curve is the entropy of the Horowitz-Polchinski  as a function of the temperature. We see that $S/S_{\textrm{ext}}$ goes to one in the extremal limit. The precise shape of the curves depends on the values of the charges and $\ell_s$ in Planck units. In the microcanonical ensemble, where we vary the mass, we expect the two curves to be joined together near where their extrapolations cross.}
\label{fig:QL=0new}
\end{figure}

We note that the low temperature thermodynamics is qualitatively different from what one would expect if there were an AdS$_2$ region in the geometry. In Section~\ref{sec:canitbe} we will argue that the fact that the low energy physics is not given by AdS$_2$ could be argued from more general principle. 
In Figure~\ref{fig:QL=0new} we illustrate that after considering the charged Horowitz-Polchinski solution, the naive extremal limit in Figure~\ref{fig:QL=0} is modified significantly.

We close this section by mentioning that other than the thermodynamic quantities, one can also look at the 
physical size of the Horowitz-Polchinski solution, and it matches with the size of the BPS string expected from a random walk 
picture~\cite{Chen:2021dsw}, 
\begin{equation}
    \ell_{\textrm{phys}} \sim \left(nw\right)^{\frac{1}{4}} \ell_s\,.
\end{equation}

\section{Gravitational configurations that contribute to the index}\label{sec:2dernonextsusy}

We now initiate our study of the gravitational path integral that evaluates the index of two-charge states. We first construct supersymmetric solutions by taking a non-trivial limit of the two-charge black hole analyzed in the previous section. The situation is more subtle than its `big black hole' counterpart and we emphasize some puzzling features. Those features will go away later when we include higher derivative corrections.

\subsection{Imposing supersymmetry}\label{sec:impsusy}

We want to construct  supersymmetric finite temperature black hole solutions that contribute to the gravitational path integral for the index. 
We actually need to consider the helicity supertrace, 
but one can simply absorb the fermion zero modes and we will not worry about this here. 
% but for now we will not worry about such details since they are not important at the classical level. 
Imposing supersymmetry at the boundary (i.e.~the asymptotic region of flat space) requires fermions to be periodic. As explained in \cite{Iliesiu:2021are} this constrains the angular velocity $\Omega$. Supersymmetry in the bulk requires the existence of a globally defined Killing spinor. For the black hole geometries presented in Section~\ref{sec:SBHthermo} this constrains the mass parameter $M$. Explicitly these two conditions are given by
\bea
&&{\rm Boundary~Supersymmetry:}~~~~\,\beta \Omega \= 2\pi \i \, ~{\rm mod}~4\pi \mathbb{Z}\, ,\nonumber\\
&&{\rm Bulk~Supersymmetry:} ~~~~~~~~~~~~M \= Q_R/\sqrt{2}.\nonumber
\ea
The first condition fixes the fermion periodicity conditions because this choice of angular velocity amounts to an insertion of $e^{\beta \Omega J} \to e^{ 2\pi \i J}= (-1)^{\sf F} $. Since all excitations have integral or half-integral angular momentum this condition defines the angular velocity only up to a shift multiple of $4\pi/\beta$. Bulk and boundary supersymmetry are in principle independent. For example, in the case of Reissner-Nordstr\"{o}m, if $\beta \Omega = 2\pi \i$ then this implies that $M=Q$ and a globally defined Killing spinor exists, but if $\beta \Omega = 6\pi \i$ or any other odd multiple of $2\pi \i$, then $M\neq Q$ and supersymmetry is not preserved in the bulk. One unexpected feature of the two-charge system is that even for $\beta \Omega = 2\pi \i$ bulk supersymmetry is not guaranteed. 

The discussion here will assume that in the presence of $(-1)^{\sf F}$ the gravitational path integral is dominated by a black hole-like saddle. Can there be another saddle similar to the Horowitz-Polchinski one that contributes to the index? One option, since the thermal circle is non-contractible, is to take the Horowitz-Polchinski solution of the previous section and impose fermions to be periodic. The problem with this proposal is that the solution is not supersymmetric and therefore does not contribute to the index due to fermion zero-modes.\footnote{Even if it did, it produces a contribution that depends on temperature in the free energy, inconsistent with the behavior of an index.} Note also that the Horowitz-Polchinski solution also breaks down before we reach exact extremality due to quantum corrections. Even if this had worked, the problem is that the equations of motion that Horowitz-Polchinski solves are not correct in the presence of $(-1)^{\sf F}$. At least in flat space, periodic fermions lead to the standard GSO projection and therefore the light winding mode that condenses in the Horowitz-Polchinski solution is not part of the low energy spectrum~\cite{Atick:1988si}. It is therefore not clear which equations we are supposed to solve to find such a solution. It is possible that by turning on an imaginary angular potential, similar to the black hole case, one can evade the flat space GSO projection argument and arrive at a supersymmetric solution, but we are not sure how to construct such a solution explicitly (see related discussion in \cite{Ceplak:2023afb}). We should emphasize that even if such a solution was found and produces the correct index, it would not invalidate our discussion below. On the contrary, it would be in harmony with the conjecture that there is no phase transition between the Horowitz-Polchinski solution and the black hole in the heterotic theory. 
Since the supergravity sector is part of the low energy spectrum when fermions are periodic, and since there are black hole solutions in this sector, this will be our focus in the rest of the paper.

\medskip

After these introductory remarks let us impose $\beta \Omega = 2\pi \i$ and see what the implications are for the two-charge black hole. Since the angular velocity is imaginary it is convenient to introduce $a = \i {\sf a}$, so ${\sf a}$ will then be real. The condition becomes then
\beq
\frac{\beta \Omega}{2\pi \i} \= \frac{{\sf a}}{\sqrt{m^2+{\sf a}^2}}\=1\,.
\eeq
It is convenient sometimes to replace the variables $(m,{\sf a})$ by $(m, x = {\sf a}/m)$. Then the boundary supersymmetry condition becomes $ x = \sqrt{1+x^2}$, and the only solution is to take $x\to \infty$. We are working in the thermal ensemble and the black hole temperature, in the limit $x={\sf a}/m\to \infty$, becomes 
\beq\label{betaaftersusy}
\beta \= \frac{2\pi m (\cosh \upalpha + \cosh \upbeta)(m+\sqrt{m^2+{\sf a}^2})}{\sqrt{m^2+{\sf a}^2}} \Big|_{{\sf a}/m\to\infty} \= 2 \pi m (\cosh \upalpha + \cosh \upbeta) \,.
\eeq
Since the temperature is a free parameter set by our choice of boundary conditions at asymptotic flat space, the right hand side of this expression should remain finite. We analyze below two ways to implement this condition.
\smallskip 

\paragraph{Setting ${\sf a} \to \infty$ and $m$ finite} As a first possibility we can keep $m$, $\upalpha$ and $\upbeta$ finite, with one combination of them set by the temperature, and send ${\sf a}\to \infty$ to implement boundary supersymmetry. We will see now this option is ruled out after imposing bulk supersymmetry. For $m$ finite, combining \eqref{MJ} with \eqref{QLQR} one can show that setting $M^2 = Q_R^2 /2 $ implies the following relation between the parameters $\upalpha$ and $\upbeta$
 \beq
(1+\cosh\upalpha\cosh \upbeta)^2 \= \cosh^2 \upalpha \sinh^2 \upbeta \,.
\eeq
 There is no real solution of this equation for finite $\upalpha$ and $\upbeta$, since the constraint implies $e^{\upalpha+\upbeta}<0$. It is not clear that the black hole solves the correct equations of motion since for complex values of $\upalpha$ and $\upbeta$ the transformation used to generate the solution is not in $\OO(7,23)$ any longer. There is one more problem with this scenario that is worth mentioning. We evaluate the area of the horizon, given in \eqref{eq:area}, in the large ${\sf a}$ limit 
\beq
A \to 4\pi m (\cosh\upalpha + \cosh \upbeta) m x  \quad \Rightarrow \quad A \= 2 \beta m x \,.
\eeq
We see that at fixed temperature and keeping $m$ finite, the area is $A \propto {\sf a} \to \infty$. Now instead of vanishing, the area actually diverges! It is not obvious that this feature is necessarily problematic, as explained in \cite{Chen:2023mbc}, but we would like to avoid it if possible since the interpretation of such a saddle is unclear. For these reasons we will not consider this direction any further.
\smallskip 

\paragraph{Setting $m \to 0$ and ${\sf a}$ finite} Another way of implementing the bulk supersymmetry condition is to keep ${\sf a}$ finite while taking the mass parameter $m$ to zero. This is the regime considered in \cite{Sen:1994eb}, in the absence of the $(-1)^{\sf F}$ insertion. Looking at the expression for the temperature given in \eqref{betaaftersusy} we see that this limit requires taking $(\cosh \upalpha + \cosh \upbeta) \to \infty$. How this limit is taken is fixed by bulk supersymmetry. For this purpose let us take the $m\to 0$ limit of the charges 
\beq\label{eq:Qsusy}
\vec{Q}_L \= \frac{\beta}{2\pi \sqrt{2}} \frac{\sinh \upalpha \cosh\upbeta}{\cosh \upalpha + \cosh \upbeta} \, \vec{n}\,, \qquad \vec{Q}_R \= \frac{\beta}{2\pi \sqrt{2}} \frac{\sinh \upbeta \cosh \upalpha }{\cosh\upalpha + \cosh\upbeta} \, \vec{p} \,.
\eeq
and similarly the mass
\beq
M \= \frac{\beta}{4\pi} \frac{1+\cosh \upalpha \cosh \upbeta}{\cosh\upalpha + \cosh \upbeta} \,.
\eeq
One can check that $\upbeta \to \infty$ while keeping $\upalpha$ finite does provide a solution of $M^2 = Q_R^2 / 2$, as pointed out in \cite{Sen:1994eb}. To guarantee both boundary and bulk supersymmetry, as well as a finite temperature, we then take
\beq
m\to0, \quad \upbeta \to \infty \,, \qquad {\rm such~that} \qquad m \cosh\upbeta \to m_0 \,,
\eeq
where we introduced the finite parameter $m_0$. In terms only of parameters that are finite in the limit, the temperature and charges are
\beq
\beta \= 2 \pi m_0\,, \qquad Q_L \= \frac{m_0}{\sqrt{2}} \sinh \upalpha \,, \qquad Q_R \= \frac{m_0}{\sqrt{2}} \cosh \upalpha \,.
\eeq
We can use the second and third equations to solve for $m_0$ and $\upalpha$ in terms of the charges. Explicitly, we obtain
\beq
m_0 \= \sqrt{2(Q_R^2 - Q_L^2)}\,, \qquad 
\tanh \upalpha \= \frac{Q_L}{Q_R}\,.
\eeq
Using these values, we see that the temperature is fixed in terms of the charges! Moreover, the parameter ${\sf a}$ is undetermined. We will comment more on this later. This solution only exists if $Q_R>Q_L$. No supersymmetric solution of this kind exists if $Q_L>Q_R$ instead.\footnote{What happens with the supergravity solution when $Q_L=Q_R$? In this case $\upalpha = \upbeta$ and bulk supersymmetry demands that $\upalpha=\upbeta \to \infty$. But it is evident from \eqref{eq:Qsusy} that in this case keeping the temperature finite leads to $Q_L,Q_R \to \infty$, therefore no solution exists with finite $Q_L=Q_R$. Another option is to take $\beta \to 0$ such that $\beta e^{\upbeta} $ is finite. This implies now that $m \sim e^{- 2\upbeta}$, vanishing faster than before. To keep the black hole area finite we need to send ${\sf a}\to \infty$. Since its not clear whether this is a physical solution or not we will not consider it further and restrict ourselves to $Q_R>Q_L$.} This is consistent with the spectrum of heterotic strings compactified on a torus, and consistent with the spectrum of Lorentzian black holes in the absence of the $(-1)^{\sf F}$ insertion discussed in Section~\ref{sec:SBHthermo}. Finally, the chemical potentials of the left and right charges have magnitudes $\mu_L=0 $ and $\mu_R=1/\sqrt{2}$, independent of the parameters $m_0$ and $\upalpha$.

\subsection{Explicit form of the solution}

Now that we have determined the scaling of all parameters in the supersymmetric limit we can write down explicitly how the solution looks like. 
The same solution was written down in \cite{Horowitz:1995tm, Dabholkar:1995nc} although it was interpreted differently. 
In the supersymmetric limit the metric simplifies slightly, 
\bea
e^{-\Phi} \, \d s^2 &\=&\Delta^{\frac{1}{2}} \Big\{ \frac{r^2 -{\sf a}^2 \cos^2 \theta}{\Delta} \d t_E^2 + \frac{\d r^2}{r^2 -{\sf a}^2} + \d \theta^2  \nonumber\\
&& \hspace{-0.4cm}+ \frac{\sin^2 \theta}{\Delta} [ \Delta - {\sf a}^2 \sin^2\theta(r^2 -{\sf a}^2 \cos^2 \theta+ 2 m_0 r \cosh \upalpha)] \,\d \phi^2 -  \frac{2 m_0 r {\sf a} \sin^2 \theta }{\Delta}\, \d t_E \d \phi \Big\}\,,\nonumber\\
\label{metricsentcbh}
\ea
where the supersymmetric limit of the function $\Delta$ becomes
\beq
\Delta \= (r^2 -{\sf a}^2 \cos^2 \theta)^2 + 2 m_0 r ( r^2 - {\sf a}^2 \cos^2 \theta) \cosh\upalpha + m_0^2 r^2 \,.
\eeq
Here, we remind the reader that $\d s^2$ is defined in the string frame. 
The dilaton is still given by $\Phi = \log (r^2 -{\sf a}^2 \cos^2 \theta)/\Delta^{1/2}$. The solution is real in Euclidean space. 
% This is a general feature when evaluating supersymmetric indices with the gravitational path integral, although it does not always hold. 
From the second term in the first line we see that the horizon is located at
\beq
r_+ \= {\sf a}\,,
\eeq
giving a geometric interpretation to ${\sf a}$. 
One can verify directly with the supersymmetric solution that smoothness of the metric at the horizon imposes values of the temperature and angular velocity consistent with the previous analysis, 
namely $\beta = 2\pi m_0$ and $\Omega = \i/m_0$, such that $\beta \Omega = 2\pi \i$. 
%This is valid everywhere 
In fact, the solution is smooth everywhere on the sphere except at the poles, which we comment on later. 
The area of the horizon in Einstein frame is
\beq
A \= \int \d \theta \d \phi \sqrt{g^E_{\theta\theta} g^E_{\phi\phi} } \=  \int \d \theta \d \phi \,m_0 {\sf a} \sin \theta  \= 4\pi m_0 {\sf a} \,,
\eeq
which is finite and non-vanishing! The rotation necessary to implement the $(-1)^{\sf F}$ has turned the small black hole (at extremality) into a macroscopic one at finite temperature for the index. This seems to suggest we are now able to use the gravitational path integral to evaluate the index of the two-charge state avoiding small black holes, but this is too naive. Even though the solution has a macroscopic horizon, the black hole geometry is actually singular at the poles of the sphere $\theta=0,\pi$ so the rotation alone has not fixed all issues. This is clear when looking in the Einstein frame metric which is given by the string frame one multiplied by $e^{\Phi}$. Near the horizon 
\beq
e^{\Phi}\big|_{r={\sf a}} \= \Delta^{-\frac{1}{2}}{\sf a}^2 \sin^2\theta \,,
\eeq
and $\Delta|_{r={\sf a}}$ is a non-vanishing finite function of $\theta$. 
Therefore we see that $e^{\Phi}$ vanishes at the two poles of the sphere. 
A more detailed analysis shows that the string frame metric is singular, while the vanishing of $e^{\Phi}$ 
at the poles makes the Einstein frame metric smooth,  
although higher derivative corrections in the Einstein frame can be large due to dilaton diverging. 

The solution is specified by the gauge potentials besides the metric. These also have a finite limit in the supersymmetric limit given by
\bea
\vec{A}_L &\=& - \frac{\vec{n}}{\sqrt{2}} \frac{m_0 r \sinh \upalpha (r^2 -{\sf a}^2 \cos^2 \theta)}{\Delta} \d t-\i \frac{\vec{n}}{\sqrt{2}} \frac{ m_0^2 r^2 {\sf a} \sinh 2\upalpha \sin^2 \theta  }{2\Delta} \d \phi \,,\nn
\vec{A}_R &\=& - \frac{\vec{p}}{\sqrt{2}} \frac{m_0 r  ((r^2 -{\sf a}^2 \cos^2 \theta)\cosh\upalpha + m_0 r)}{\Delta} \d t \nonumber\\
&&~~~~~+ \i \frac{\vec{p}}{\sqrt{2}} \frac{ m_0 r {\sf a} \sin^2 \theta  (r^2 -{\sf a}^2 \cos^2 \theta + m_0 r \cosh\upalpha )}{\Delta} \d \phi\,,\nn
B&\=&-\i \frac{m_0 r {\sf a} \sin^2 \theta}{\Delta}( r^2 -{\sf a}^2 \cos^2 \theta + m_0 r \cosh\upalpha ) \,\d t  \d \phi \,.
\ea
These gauge potentials are smooth and one can extract from them the chemical potentials, verifying that $\mu_L=0$ and $\mu_R = 1/\sqrt{2}$. Finally, the functions appearing in the moduli scalar profiles $M$ also simplify
\beq
{\sf P} \= \frac{2 m_0^2 r^2 \sinh^2 \upalpha }{\Delta}\,, 
\qquad {\sf Q} \= -\frac{2 m_0 r \sinh \upalpha }{\Delta}(r^2 -{\sf a}^2 \cos^2 \theta + m_0 r \cosh\upalpha ) \,.
\eeq
These functions are smooth and finite at the horizon.

In general, the Euclidean action of the two-charge heterotic black hole is given by $I_{\rm on-shell}= \beta M - A/4 - \beta \Omega J$ and comes solely from the boundary terms in the action. 
This is the result in an ensemble of fixed electric charges. If we have a supersymmetric solution then the first term is $\beta Q_R^2/2$. In the rest of the paper we will always include an appropriate boundary term that has the effect of shifting the zero-point energy such that $I \to I - \beta Q_R^2/2$.
After this modification, $-I_{\rm on-shell}$ simplifies to $A/4 + \beta \Omega J$. 
We can show now that whenever a supersymmetric saddle exists, then these two terms cancel each other and the classical two-derivative contribution to the on-shell action, and therefore to~$\log Z_{\rm index}$ vanishes. 
The area and the angular momentum can be written as
\beq
\frac{A}{4} \= \frac{ \beta {\sf a}}{2}\,, 
\qquad J \=\i \frac{\beta {\sf a}}{4\pi} \,.
\eeq
We already discussed the area. The angular momentum can be read off directly from the supersymmetric solution or taking the appropriate limit of \eqref{MJ}. Using that $\beta \Omega = 2\pi \i$ we get the desired cancellation
\beq
\frac{A}{4} + \beta \Omega J \= 0\,, \qquad 
\Rightarrow \qquad \log Z_{\rm index} \=- I_{\rm on-shell} \= 0 \,.
\eeq
We found a black hole with a macroscopic area that nevertheless gives a vanishing contribution to the index to leading order. This is the desired conclusion, since a non-zero action at the two-derivative level would, by dimensional analysis, grow too fast with the charges and would not reproduce the microscopic index. 

\subsection{Summary of properties}\label{sec:susybh2derprop}

To recap, we have found a supersymmetric finite temperature black hole with two-charges that evaluates the index avoiding the small black hole regime. 
Even after including the rotation necessary to implement the $(-1)^{\sf F}$ some puzzles remain to be resolved. 
We finish this section by summarizing and expanding on some of them. In the next sections we will resolve them by including higher derivative corrections. 

\begin{enumerate}
    \item The solution we found has a curvature singularity at the poles of the $S^2$ horizon $r=r_+={\sf a}$ at $\theta=0$ and $\pi$. This forces us to include higher derivative corrections even though the horizon is macroscopic.
    \item For a given set of charges, the solution only exists for a unique value of the temperature. This is surpring since charges and temperatures can be independently determined by boundary conditions at infinity. After taking the supersymmetric limit we found $$\beta \= 2\pi \sqrt{2(Q_R^2-Q_L^2)} \,.$$ For comparison, in the Reissner-Nordstr\"{o}m case analyzed in \cite{Iliesiu:2021are} the inverse temperature is $\beta = \pi Q(Q+2{\sf a})/{\sf a}$, while $r_+ = Q+{\sf a}$. A solution therefore exists for any temperature and the horizon radius depends on its value. For the case considered here, the singular feature is that the dependence on ${\sf a}$ disappeared completely from $\beta$, making it a function only of the charge. What solution contributes to the gravitational path integral when $\beta \neq 2\pi \sqrt{2(Q_R^2-Q_L^2)}$?
    \item In the Reissner-Nordstr\"{o}m case quoted in the previous item, not only is the temperature a tunable parameter but it also determines a unique value of ${\sf a}$. In the two-charge case another consequence of the fact that ${\sf a}$ drops out of the temperature\footnote{The fact that ${\sf a}$ drops out, and the temperature is fixed in terms of the charges, is special to compactifications to four dimensions. Similar black hole solutions in toroidal compactifications to $D>4$ dimensions do not have this problem \cite{Chowdhury:2024ngg}. The reason we insist on working in $D=4$ is the fact that supersymmetric higher derivative solutions are known and we use them in Section~\ref{sec:higherdersol}.  } is that there is a moduli space of solutions labeled by ${\sf a}$ from $0$ to $\infty$. All values of ${\sf a}$ contribute equally and have to be integrated over. This is a disturbing state of affairs and we will see that higher derivative corrections will determine a unique value of ${\sf a}$. 
\end{enumerate}

Before addressing the effect of higher derivative corrections, we will show in the next section that these 
features are common to a class of two-charge black holes in $\mathcal{N}\geq 2$ supergravity.

\section{The index and new attractor mechanism in 4d $\CN=2$ supergravity \label{sec:4dsugra}}

In this section we move to the~$\CN=4$ four-dimensional effective theory that describes 
the compactification of the heterotic string on~$T^6$. From the string-string 
duality, this is equivalent to Type II theory on~$K3 \times T^2$. More generally, 
we can consider four-dimensional~$\CN=2$ theories that arise from the 
compactification of Type II on a Calabi-Yau 3-fold.
These theories are described by~$\CN=2$ supergravity coupled to vector and hyper multiplets. 
The fields that participate in the black hole solutions live in the graviton multiplet and the vector multiplet, and we focus on this sector. 
In this section we discuss the theory with a two-derivative action, and in the next section we consider higher-derivative corrections.

The conformal supergravity formalism is an elegant and powerful formalism to describe the theory of $\CN=2$ supergravity coupled to a number of vector multiplets~\cite{deWit:1979dzm, deWit:1980lyi}. In this formalism, local conformal invariance is a gauge symmetry of the theory. 
Accordingly, each field is assigned a conformal or Weyl weight. 
Eventually, the conformal symmetry is gauge-fixed by assigning a dimensionful number like~$m_\text{Pl}$ 
to a field with non-zero Weyl weight. This gauge-fixing leads to Poincar\'e supergravity.

The bosonic physical field content of each vector multiplet consists of 
a vector field~$A^I$ and a complex scalar~$X^I$, $I=0,1,\dots, \nv$. 
The scalars have Weyl weight~1.
Solutions of the theory carry electric and magnetic charges~$(Q_I, P^I)$ under these gauge fields. 
The two-derivative action is completely specified by the prepotential~$F(X)$,
which is a homogeneous function of degree two and has Weyl weight~2. 
An important role is played 
by the first derivatives~$F_I=\p F/\p X^I$ which have Weyl weight~1.
At the two-derivative level the equations of motion of the theory have~$Sp(2\nv+2;\IR)$ symmetry,  
under which the \emph{charge vector}~$\Gamma=(P^I,Q_I)$ and the 
\emph{period vector}~$\Omega_{\text{hol}}=(X^I,F_I)$ 
transform as~$2(\nv+1)$-dimensional vectors.

Two important invariants of the symplectic symmetry are\footnote{Here, and below, the symplectic product of 
two vectors~$A=(\wt{A}^I,A_I)$ and~$B =(\wt{B}^I, B_I)$ is given 
by~$\langle A ,  B \rangle = \wt A^I  B_I - \wt B^I A_I$.}  
the generalized K\"ahler potential (of Weyl weight~2) 
\be \label{defK}
\rme^{-\CK(X,\overline{X})} 
\= \i \bigl( X^I \overline{F}_I -\overline{X}^I F_I  \bigr) \=  \i \intprod{\Omega_{\text{hol}}}{\overline{\Omega}_{\text{hol}}} \,,
\ee
% I kept this symplectic product notation because of the final equation for the entropy which is nicely written in terms of it. 
and the central charge function (of Weyl weight~0) , 
\be \label{defZ}
Z(\Gamma;\Omega) \= 
\intprod{\Gamma}{\Omega}
\= \rme^{\half\CK(X,\overline{X})}\,(P^I F_I-Q_I X^I ) \,.
\ee
The conjugate central function~$\overline{Z}(\Gamma;\Omega) \= Z(\Gamma;\overline{\Omega})$ is also used below.

\subsection{$\frac12$-BPS solutions of~$\CN=2$ supergravity}

The general $\frac12$-BPS stationary solution of this supergravity theory is described as 
follows~\cite{Behrndt:1997ny, LopesCardoso:2000qm}.  
The input to the solution 
the vector~$\Hv(\xvec) =\bigl(\wt H^I(\xvec),H_I(\xvec) \bigr)$ 
of harmonic functions on base space. 
These functions are sources of electric and magnetic fields. 

The metric takes the form 
\be \label{IWPmetric}
\dd s^2 \=  \rme^{2U} (\dd t_E+i\omega)^2 +  \rme^{-2U} \dd x^m \dd x^m \,,
\ee
where~$U(\xvec)$ is a function of the base space coordinates~$x^m$, $m=1,2,3$, 
and~$\omega=\omega_m \dd x^m$ is a one-form on the base space that governs the
angular rotation of the solution. 
In order to have asymptotically flat space, one has~$\rme^{-2U(\infty)}=1$. 
Note that we have directly described the Euclidean solution with compact time coordinate~$t_E$. 
The warping of the base space and the connection~$\omega$ in~\eqref{IWPmetric} 
are given by 
 \be  \label{Uomgeneral}
 \rme^{-2U} \= 
    \ii \bigl(\CY^I \, \overline{\CG}_{I} - \overline{\CY}^I \, \CG_{I} \bigr)    \,,
\qquad \star \, \dd \, \omega \= i \langle \dd H , H \rangle \,,
% \ve_{mnp}\, \p_n \omega_p  \=  H_I \p_m \wt H^I - \wt H^I \p_m H_I  \,.
\ee
where the weight-0 normalized fields~$( \CY^I ,  \CG_I )$ are given by 
\be \label{defYG}
\bigl(\CY^I, \CG_I \bigr) \= e^{\frac12 \CK(X, \overline{X})} \, \overline{Z}(\Hv;\Omega) \, \bigl(X^I, F_I \bigr) \,
\ee
and similarly~$\overline{\CY}^I$, $\overline{\CG}_I$ by the complex conjugate equations. The  scalar fields obey the \emph{generalized stabilization equations}\footnote{Note that the signs in these equations are opposite to the one in~\cite{Boruch:2023gfn}.} 
\be \label{genstab}
\CY^I - \overline{\CY}^I \= \ii \wt H^I \,, \qquad 
\CG_I -  \overline{\CG}_I  \= \ii H_I  \,,
\ee
The above equations of supersymmetry for  the~$(\CY^I, \CG_I )$ fields can be presented in 
terms of a prepotential function~$\CG(\CY)$ with~$\CG_I = \partial \CG/\partial \CY^I$, 
where~$\CG$ is the same mathematical function as~$F$.\footnote{One should note that the 
rescaling between~$X^I$ and~$\CY^I$ is not holomorphic, but one can still apply the concept of 
holomorphic and anti-holomorphic to the~$\CY$ variables directly. 
Indeed, we encounter holomorphic (and anti-holomorphic) functions of~$\CY^I$ in the discussion of the entropy.}

Besides the metric and the scalars, the solution is specified by the $\nv+1$ vector multiplet ${\rm U}(1)$ gauge fields. 
These are also determined by the harmonic functions in a specific way, presented for example in equation (3.21) of~\cite{Boruch:2023gfn}.

\subsection{Euclidean black holes and  near-horizon behavior \label{sec:newattr}}

In this subsection we review the ``new attractor mechanism'' that was found in~\cite{Boruch:2023gfn}.
Here we present all the equations in coordinate language. We refer the reader 
to~\cite{Boruch:2023gfn} for details of the  coordinate-invariant equations. 

\bigskip

\ndt {\bf Extremal black holes}  Supersymmetric extremal solutions are obtained by 
choosing the harmonic functions in the above discussion 
to have electric and magnetic sources at one point, say the origin, i.e.,
\be \label{Hextr}
H(\xvec)\=h+\frac{\Gamma}{r}\,,  \qquad r=|\xvec | \,.
\ee
The equations~\eqref{Uomgeneral} show that the solution is spherically symmetric and static, i.e.~$\omega=0$ in~\eqref{IWPmetric}. 
The vector~$h\equiv H(\infty)$ is the asymptotic value of sources.  
The generalized stabilization equations~\eqref{genstab} imply that the asymptotic 
behavior of the scalars as~$r \to 0$ is
\be
\CY^I \, \sim \, \frac{Y^I_*}{r}  \,, \qquad \CG_I \, \sim \, \frac{G_{I*}}{r} \,, 
\ee
where the constant values~$Y^I_*, G_{I*}$ (and therefore $Z_*$) are determined by the \emph{extremal attractor equations}
\be \label{attractoreqns}
Y^I_*  - \overline{Y}^I_*  \= \ii P^I \,, \qquad 
G_{I*}  - \overline{G}_{I*}  \= \ii Q_I \,.
\ee
The behavior of the metric as~$r \to 0$ is
\be \label{USrel}
 \rme^{-2U} \, \sim \, 
    \frac{ \ii \bigl(Y^I_* \, \overline{G}_{I*} - \overline{Y}^I_* \, G_{I*} \bigr) }{r^2}  \,.
\ee
From the form of the metric~\eqref{IWPmetric} we see that the Bekenstein-Hawking entropy of the black hole 
is given by~$S_\text{BH}=\pi  \i \bigl(Y^I_* \, \overline{G}_{I*} - \overline{Y}^I_* \, G_{I*} \bigr)$.

\vskip0.4cm

\ndt {\bf Non-extremal supersymmetric solutions} 
%The idea behind the non-extremal supersymmetric solutions is to
In order to obtain non-extremal solutions relevant for the index, we replace the double pole
behavior~\eqref{USrel} of the extremal metric by two single poles. 
Accordingly, we split the sources into a pair of sources at two different points called the north and south pole.
The total charge is split as
\be 
\label{chargesplit}
\gamma_N \= \frac{1}2 \bigl(P^I+\i n^I, Q_I + \i m_I \bigr)\,, \qquad 
\gamma_S \= \frac{1}2 \bigl(P^I-\i n^I, Q_I - \i m_I  \bigr) \,, \qquad 
\Gamma\=\gamma_N \,+\, \gamma_S\,,
\ee 
and the corresponding harmonic functions are given by  
\be \label{HNSsplit}
\Hv(\xvec) \= h + \frac{\gamma_N}{|\xvec-\xvec_N|} + \frac{{\gamma}_S}{|\xvec-{\xvec}_S|}\,. 
\ee
Compared to the extremal solution we have introduced electric and magnetic dipole charges $(m_I,n^I)$ 
and the distance~$|\xvec_N-\xvec_S|$ between the north and south poles. 
The equation for~$\omega$ in~\eqref{Uomgeneral} shows that the metric necessarily has angular momentum. 
All these parameters of the solution are determined by the \emph{new attractor mechanism} \cite{Boruch:2023gfn} that we now recall.

The main point is that the spinning non-extremal solution near one of the poles has a simple scaling behavior, 
and the values of the scalars at the poles are therefore determined completely by the charges, 
similar to the extremal solution. 
There is, however, an important difference in that now we only have simple poles.
As we briefly review below, one of the consequences is that the solution is necessarily complex.

The full solution has a symmetry which exchanges the north with the south pole and the scalars with their complex conjugates. 
Therefore we focus on the north pole and let the parameter~$\rho  = |\xvec - \xvec_N|$ be very small.  
From~\eqref{HNSsplit} we have, as~$\rho \to 0$, 
\be
H(\xvec) \; \sim \; \frac{\gamma_N}{\rho} + 
H^{(0)} + \text{O}(\rho) \,,  \qquad 
H^{(0)} \= h + \frac{\gamma_S}{|\xvec_N-\xvec_S|} \,.
\ee
The generalized stabilization equations~\eqref{genstab} then imply that  
\be \label{YGLaurent}
\CY^I \; \sim \; \frac{Y^I_N}{\rho} + Y^{I(0)} + \text{O}(\rho) \,, \qquad 
\CG_I \; \sim \; \frac{G_{NI}}{\rho} + G^{(0)}_I + \text{O}(\rho) \,, \qquad 
\ee
and similarly for the conjugate scalars, 
with 
\be \label{YYbarNeqns}
\bigl( Y^I_N - \overline Y^I_N  \,, \, G_{IN} - \overline G_{IN} \bigr) \= \ii \gamma_N \,, \qquad 
\bigl( Y^{I(0)} - \overline Y^{I(0)} \,,\, G^{(0)}_{I} - \overline G^{(0)}_{I} \bigr) \= \ii  H^{(0)} \,.
\ee
Note the Laurent expansion~\eqref{YGLaurent} combined with the definition of the scalar fields~\eqref{defYG}
and the homogeneity property of the prepotential leads to the following identity that we use below,
\be \label{GIJid}
G^{(0)}_I  \= \frac{\partial \CG_{I}}{\partial \CY^J}(\xvec_N) \, Y^{J(0)} \quad \Rightarrow \quad 
Y^I_N \, G^{(0)}_I  \= G_{JN} \,  Y^{J(0)} \,.
\ee

\medskip

Now, the equation~\eqref{Uomgeneral} for the metric implies that 
 \be  
 \rme^{-2U} \; \sim \; 
    \ii \Bigl(\frac{Y^I_N}{\rho} + Y^{I(0)}  \Bigr) \Bigl(  \frac{\overline G_{IN}}{\rho} + \overline G^{(0)}_I   \Bigr)  
 - \ii  \Bigl(\frac{\overline Y^I_N}{\rho} + \overline Y^{I(0)} \Bigr) \Bigl(\frac{G_{IN}}{\rho} + G^{(0)}_I \Bigr) + \text{O}(\rho)    \,.
\ee
Demanding a simple pole at~$\xvec_N$ implies that
\be \label{YNvals}
\bigl(\overline Y^I_N \,, \overline G_{IN} \bigr) \= 0 \quad \Rightarrow \quad 
\bigl( Y^I_N  \,, \, G_{IN}  \bigr) \= \ii \gamma_N \,,
\ee
where we use the first equation of~\eqref{YYbarNeqns} to obtain the second equality.  
The real part of these equations,
\be
 Y^I_N - (Y^I_N)^*  \= \ii P^I \,, \qquad 
G_{IN} - (G_{IN})^*   \= \ii Q_I \,,
\ee
are recognized as the extremal attractor equations~\eqref{attractoreqns}.
Assuming a unique solution, we obtain~$(Y^I_N, G_{IN}) = (Y^I_*, G_{I*})$. 
The scalars at the north pole are thus fixed by the solution to the extremal attractor equations. 
Further, the dipole charges can then be read off to be 
\be
-n^I \= Y^I_N + (Y^I_N)^* \,, \qquad - m_I \= G_{IN} + (G_{IN})^* \,.
\ee
The metric now only has a simple pole,
\be  \label{e2UNP}
\begin{split}
 \rme^{-2U} & \; \sim \; 
    \frac{\ii}{\rho} \Bigl( Y^I_N \,  \overline G^{(0)}_I  - G_{IN} \, \overline Y^{I(0)}  \Bigr) + \text{O}(1)   \,,\\
&  \= -  \frac{1}{\rho} \Bigl( Y^I_N \,  H^{(0)}_I  -  G_{IN} \, \wt H^{I(0)} \Bigr) 
 +  \frac{\ii}{\rho} \Bigl( Y^I_N \,  G^{(0)}_I  -  G_{IN} \,  Y^{I(0)}  \Bigr) + \text{O}(1)   \,, \\
& \= -  \frac{1}{\rho} \Bigl( Y^I_N \,  H^{(0)}_I  - G_{IN} \, \wt H^{I(0)}   \Bigr) + \text{O}(1) \,.
\end{split}
\ee
Here, we use the second equation in~\eqref{YYbarNeqns} to reach the second line, and the identity~\eqref{GIJid}
in order to obtain the final equality.

Finally, the behavior of the connection~$\omega$ near the poles is given by expanding the equation~\eqref{Uomgeneral}. 
Using~\eqref{YGLaurent} we obtain 
\be \label{Romegaval}
\star \, \dd \, \omega \=  \frac{\wh n}{\rho^2} \langle \gamma_N, H^{(0)}\rangle + \text{O}(1/\rho)\,,
\ee
where~$\wh n = (\xvec-\xvec_N)/|\xvec-\xvec_N|$, the three-dimensional unit vector with the north pole as origin. 
This equation implies that there is a three-dimensional Dirac string associated with~$\omega$ 
emanating from the north pole. There is a similar condition at the south pole. 
The condition of smoothness implies that this Dirac string should end smoothly at the poles. 
This regularity condition leads to (see the discussion in \cite{Hartle:1972ya,Whitt:1984wk} for more details)
\be \label{gensmoothcond}
\frac{\i \langle \gamma_N , {\gamma}_S \rangle}{|\xvec_N - {\xvec}_S|} 
\; + \; \i \langle \gamma_N ,h \rangle \= \frac{\beta}{4\pi} \, . 
\ee
This equation fixes the distance between the north and south poles in terms of the temperature, 
charges at the poles, and the asymptotic values of the moduli. 
As~$\beta \rightarrow \infty$, we obtain the attractor solution with~$\Omega \rightarrow 0$,~$|\xvec_N - {\xvec}_S| \to 0$,
and an infinite throat.

To summarize, we obtain a Euclidean metric with a cigar-like topology. The consistency of the periods of angular and time coordinates,
leads to the condition \hbox{$\beta \Omega = 2\pi \i$}. 
The values of the scalars at the tip of the cigar 
at the north and south poles are fixed by the charges in a manner similar to the extremal attractor mechanism, but 
now all the equations are split into a holomorphic piece at the north pole and an anti-holomorphic piece at the south pole. 
The holomorphic scalar fields take their attractor values at the north pole,
and the anti-holomorphic scalars vanish. 
At the south pole we have the conjugate equations, and the anti-holomorphic scalars take their attractor values and the holomorphic scalars vanish.

\subsection{Two-charge solution \label{sec:2charge}}

With this set-up, we study the behavior of these~$\frac12$-BPS rotating Euclidean solutions with two-charges. 
We begin by making some general comments that apply to any theory of $\mathcal{N}=2$ supergravity. We will make contact with the black hole studied in sections~\ref{sec:SBHthermo} and~\ref{sec:2dernonextsusy} at the end. 
The classical on-shell action of the non-extremal supersymmetric solution reviewed above is given, after removing a factor of $\beta M_{\rm BPS}$ by a boundary term, by 
% \bea
% \label{IGGbar}
% -I_{\rm on-shell} 
% &\=& \pi (q_I \CY^I + 2\i \CG) \big|_N + \pi (q_I \bar{\CY}^I -2\i \bar{\CG} )\big|_S\,,\\
% &\=& \pi \i \langle \gamma_N , \gamma_S\rangle\,,\label{IGGbar2der}
% \ea
% where $\CG$ is the same function of $\CY^I$ as $F(X^I)$. 
\bea
\label{IGGbar}
-I_{\rm on-shell} 
&\=& \pi (q_I Y^I_N + 2\i \CG(Y_N)) + \pi (q_I \overline{Y}_S^I -2\i \overline{\CG(Y_S)} )\,,\\
&\=& \pi \i \langle \gamma_N , \gamma_S\rangle\,,\label{IGGbar2der}
\ea
where $\CG$ is the same function as $F$, as discussed below~\eqref{genstab}. 
The result in the first line comes from showing that the on-shell action is a total derivative up to delta functions localized at the poles~\cite{Boruch:2023gfn}. 
Using the new attractor equations it can be put in the form of the second line\footnote{The rewriting in the second line is only valid at the two-derivative level.}. 
This quantity $\pi \i \langle \gamma_N, \gamma_S\rangle$ is \emph{not} the area of the horizon of the non-extremal solution, which is instead equal to
\beq\label{eq:arean2}
A \= \beta |\xvec_N - \xvec_S| \,.
\eeq
This can be easily derived (see~\cite{Boruch:2023gfn}) from the metric~\eqref{IWPmetric} together with the equation for $\omega$ in~\eqref{Uomgeneral} 
and we will not repeat it here. 
The on-shell action and the area only coincide at zero temperatures, such 
that~$\log Z_{\rm index}$ is proportional to the extremal area. 
This is due to the fact that these black holes satisfy the quantum statistical relation, 
i.e.~the on-shell action is equal 
to~$A/4 + 2\pi \i J$ (since $\beta \Omega = 2\pi \i$) and at low temperatures the angular momentum vanishes.

Given the discussion in the previous paragraph, in a general $\mathcal{N}=2$ theory of supergravity, solutions that have vanishing area at extremality are characterized by
\beq \label{gammaNS0}
\langle \gamma_N,\gamma_S\rangle \=0 \,.
\eeq
This condition guarantees that the extremal area of the black hole is zero at the two-derivative level, and the index of the black hole is small. 
Since $\gamma_{N/S}$ are determined in terms of the solution to the extremal attractor 
equations, this can be rephrased as a condition on the choice of charges that the black hole carries. 
Finally, as emphasized above, the condition~\eqref{gammaNS0} 
%$\langle \gamma_N,\gamma_S\rangle$ 
still allows for the 
finite temperature supersymmetric solution of \cite{Boruch:2023gfn} to have a finite size horizon with area \eqref{eq:arean2}.

For the choice of charges that satisfy~$\langle \gamma_N, \gamma_S \rangle = 0$, this might 
suggest that we have succeeded in finding a smooth solution that contributes to the gravitaitonal path integral for the index. 
However, this is too quick. Let us recall the regularity condition at the 
horizon~\eqref{gensmoothcond}. 
When charges are chosen to satisfy $\langle \gamma_N, \gamma_S \rangle = 0$, the condition degenerates to:
\beq\label{eq:tempsmallbh}
\beta \= 4 \pi \i \langle \gamma_N , h\rangle \,.
\eeq
This is problematic for the following reasons
\begin{enumerate}
    \item First, the temperature \eqref{eq:tempsmallbh} is fixed in terms of the black hole charges and asymptotic moduli, while we should be free to set them independently. 
    
    \item The second problem is that $|\xvec_N - \xvec_S|$ has disappeared completely from the equation. There is nothing fixing the separation between the two poles and therefore there is a moduli space of solutions, characterized for example by the area of the horizon. In particular $|\xvec_N - \xvec_S|=0$ is part of the moduli space which gives rise to the original singular small black hole we were trying to avoid.
\item The extremal area is also proportional to $Z_*(\Gamma) \overline{Z}_*(\Gamma)$, which implies $Z_*(\Gamma)=0$ for the small black hole. In the non-extremal solution $Z=0$ and $\bar{Z}=0$ at the north and south poles. The solution is then singular whenever $\langle \gamma_N, \gamma_S \rangle = 0$.

\end{enumerate}
These are precisely the shortcomings enumerated in Section~\ref{sec:susybh2derprop} for the two-charge black hole in the heterotic string. 
We see here all those features are universal to $\mathcal{N}=2$ supergravity with vanishing extremal area at the two-derivative level.

To provide some more intuition we will consider some special cases, and connect to the black hole in Section \ref{sec:2dernonextsusy}. 
% From here
To leading order, compactifications of type II or heterotic string theory to four dimensions can be described by the following prepotential 
% \be \label{Ftree}
% \CG(\CY) \= -\frac{1}{6}\frac{C_{IJK}\CY^I \CY^J \CY^K}{\CY^0} ,
% \ee
\be \label{Ftree}
F(X) \= -\frac{1}{6}\frac{C_{IJK}X^I X^J X^K}{X^0} \,.
\ee
The value of $C_{IJK}$ depends on the model. For example for type IIA compactified on $K3 \times T^2$ the coefficients $C_{IJK}$ are the intersection numbers of a basis 
of 4-cycles. This case is related by duality to the toroidally compactified heterotic string theory in Sections~\ref{sec:SBHthermo} and~\ref{sec:2dernonextsusy}. 
% Until here
Instead of studying this theory we will analyze a simpler version where only $X^1,X^2,X^3$ are non-zero and moreover $X^2=X^3$. 
Therefore consider a theory with $\nv =2$ and a prepotential of the form
% \be \label{Ftree}
% \CG(\CY) \= -\frac{\CY^1 (\CY^2)^2}{\CY^0} 
% \ee
\be \label{Ftreenv2}
F(X) \= -\frac{X^1 (X^2)^2}{X^0} \,.
\ee
This theory is equivalent to a sector of~\eqref{action}, as explained in~\cite{Sen:2004dp}. 
We make more comments on this below. A concrete solution with vanishing extremal area involves turning on only the following two charges
\beq
Q_I \= (Q_0 , 0,0)\,, \qquad P^I \= (0, P^1, 0) \,.
\eeq
We take $Q_0<0$ and $P^1>0$. In the context of the toroidally compactified heterotic string these charges have the interpretation of momentum $n$ and winding $w$ along a circle $S^1 \subset T^6$, namely
\beq
Q_0 \= - n\,, \qquad P^1 \= w \,.
\eeq
Moreover the interpretation of the moduli is $X^1/X^0 = \i e^{-\Phi}$ 
while $X^2/X^0$ parametrizes the volume of a cycle inside $T^6$. 
The solution to the extremal attractor equations~\eqref{attractoreqns} is given by
\beq
Y_*^I \= \Big(0,\frac{\i P^1}{2},0\Big)\,, \qquad G_{I*} \=\Big(\frac{\i Q_0}{2}  , 0, -\sqrt{P^1 |Q_0|}\Big) \,.
\eeq
This result should be surprising. Since $G_{2*} = - 2 Y^1_* Y^2_* /Y^0_*$, 
how come $X^0 =X^2=0$ and $G_{2*}$ is finite? What is the physical value of the moduli $X^2/X^0$ that describes the geometry of $T^6$? 
To see how this comes about let us turn a small magnetic charge $P^2$. 
(Physically the way this is regulated is by turning on $\wt H^2 =h^2 \neq 0$ and going slightly away from the horizon, 
keeping $P^2=0$. In practice it is easier to turn on a small $P^2$ at the attractor point, 
both lead to the same answer.) Then the attractor equation becomes
\beq
Y_*^I \= \Big(\frac{P^1 P^2}{2 \sqrt{P^1 |Q_0|}},\frac{\i P^1}{2},\frac{\i P^2}{2}\Big) \,.
\eeq
We see that as $P^2\to 0$ the ratio between $Y^2_*$ and $Y^0_*$ is finite
\beq
\frac{Y^2_*}{Y^0_*} \; \to \;  \i \sqrt{\frac{|Q_0|}{P^1}} \= \i \sqrt{\frac{n}{w}} \,.
\eeq
In the application to the heterotic string this guarantees that near the horizon (or near $\xvec_N$ and $\xvec_S$ for the non-extremal metric) the volume of $T^6$ is finite, and also explains why $G_{2*}$ is non-vanishing. The other moduli controlling the string coupling does diverge
\beq
\frac{Y^1_*}{Y^0_*} \= \i \sqrt{P^1 |Q_0|} \frac{1}{P^2} \; \to \;  \i\sqrt{nw}\, \times \infty \,.
\eeq
This divergence will be regulated by higher derivative corrections. In string theory this implies that the string coupling vanishes at the horizon (extremal) or at the poles $\xvec_{N/S}$ (non-extremal).

The new attractor equations imply that $(Y^I_*,G_{I*})$ and $(\bar{Y}^I_*, \bar{G}_{I*})$ are precisely the local charge at the north and south poles respectively
\beq
\gamma_N \= \Big( 0,  \frac{P^1}{2}, 0; \frac{Q_0}{2}  , 0 , \i \sqrt{P^1 |Q_0|} \Big),~~~\gamma_S = (\gamma_N)^* \,.
\eeq
We can replace these local charges into equation \eqref{HNSsplit}. Once these harmonic functions are known, the full non-extremal solution can be constructed. Let us assume for simplicity that the moduli at infinity are chosen such that $H_1=\widetilde{H}^0=0$ and $\widetilde{H}^2=h^2$. We also choose $h_2=0$, although this does not imply $H_2=0$ since $\gamma_{N/S}$ include a dipole in this direction. For concreteness we also pick $X^1/X^0|_{\infty} = \i 2$ and $X^2/X^0 |_{\infty} = \i$, implying $h^1=-h_0=1$ and $h^2=1/2$.\footnote{This specific choice of asymptotic moduli is done mainly to connect later to the solution in Section~\ref{sec:2dernonextsusy}. The solution can be easily generalied to arbitrary $X^{1,2}/X^0|_{\infty}$.} The emblackening factor is
\bea
e^{-2U} &\=& 2 \tilde{H}^2 \sqrt{-\widetilde{H}^1 H_0}\\
&\=& \Big\{ \Big(1 + \frac{\frac12 P^1}{|\xvec-\xvec_N|} + \frac{\frac12 P^1}{|\xvec-{\xvec}_S|}\Big)\Big(1 + \frac{\frac12 |Q_0|}{|\xvec-\xvec_N|} + \frac{\frac12 |Q_0|}{|\xvec-{\xvec}_S|}\Big) \Big\}^{\frac12} \,.
\ea
Note that $e^{-2 U} = -\i \CY^1/\CY^0=e^{-\Phi}$. Curiously this is independent of $H_2$, which carries the information of the dipole. (Of course the dipole does appear in the scalars and gauge fields.) One can readily verify that the charge vector at the poles satisfies $\langle \gamma_N, \gamma_S \rangle = 0$. All the previous considerations then apply to this black hole. A solution only exist with a unique temperature, it has a moduli space of solutions labeled by the horizon area, and the scalars are singular at the poles of the horizon.

Let us now show that this solution is precisely the same as the one analyzed in Section~\ref{sec:2dernonextsusy} 
earlier\footnote{The reader might wonder why we used two different formalism to describe the solution. 
The presentation in Section~\ref{sec:2dernonextsusy} is useful since it can be extended outside of the supersymmetric regime as in Section \ref{sec:SBHthermo}. 
The advantage of the formalism in this section is that, within supersymmetric solutions, it will allow us to incorporate higher derivative corrections, as in Section \ref{sec:higherdersol}. }. 
This has to be true since it was proven in \cite{Sen:2004dp} that the action \eqref{action} can be obtained from $\mathcal{N}=2$ supergravity with prepotential \eqref{Ftreenv2}, so we will be brief. Consider the metric given in \eqref{metricsentcbh} and change variables $(t_E, r, \theta,\phi)$ to $(t_E, x^1,x^2,x^3)$ according to
\bea
x^1&\=&\sqrt{r^2 -{\sf a}^2} \sin \theta \cos \phi\,, \\
x^2&\=&\sqrt{r^2 -{\sf a}^2} \sin \theta \sin \phi\,,\\
x^3&\=&r \cos \theta \,.
\ea
In this coordinates the metric \eqref{metricsentcbh} has the form \cite{Chowdhury:2024ngg}
\beq
\d s^2 \=  e^{2\Phi} (\d t_E + \i \omega )^2 + \d x^m \d x^m \,.
\eeq 
where $\i \omega ={\sf a} m_0 (r^2 - {\sf a}^2 \cos^2 \theta)^{-1} r \sin^2 \theta \d \phi $ and the dilaton becomes
\beq
e^{-2\Phi}\=\Big(1+ \frac{m_0 e^{\upalpha}}{2 |\xvec-\xvec_N|}+ \frac{m_0 e^{\upalpha}}{2 |\xvec-\xvec_S|}\Big)\Big(1+ \frac{m_0 e^{-\upalpha}}{2|\xvec-\xvec_N|}+ \frac{m_0 e^{-\upalpha}}{2|\xvec-\xvec_S|} \Big) \,,
\eeq
with $\xvec_N=(0,0, {\sf a} )$ and $\xvec_S= (0,0,-{\sf a} )$. This metric looks different than the one in \eqref{IWPmetric}. The solution is to recall \eqref{metricsentcbh} is given in string frame while the formalism used here for $\mathcal{N}=2$ supergravity generates an action with canonical Einstein-Hilbert term. Therefore we should translate \eqref{metricsentcbh} into Einstein frame, which gives 
\beq
e^{-\Phi}\d s^2 \= e^{\Phi} (\d t_E + \i\omega )^2 + e^{-\Phi} \d x^m \d x^m \, .
\eeq
It is possible now to put this metric in the form \eqref{IWPmetric} with $2U=\Phi$ for a suitable choice of the constants $h$. This also implies that $P^1 = m_0 e^{\upalpha}$ while $Q_0 =- m_0 e^{-\upalpha}$. This is consistent with the interpretation of $P^1$ as winding and $Q_0$ as momentum. More importantly, the separation between the poles is identified with
\beq
|\xvec_N-\xvec_S| \= 2 {\sf a} \,.
\eeq
For example, \eqref{eq:arean2} implies that the horizon metric in IWP form is $A= 2 \beta {\sf a}$ which matches the result in Section \ref{sec:2dernonextsusy}. Moreover, for our choice of $h$ and $\gamma_N$ we obtain $\i \langle \gamma_N, h\rangle = \sqrt{P^1 | Q_0|}/2 = m_0/2$ which implies from \eqref{eq:tempsmallbh} a value of the temperature $\beta = 2\pi m_0$, matching Section \ref{sec:2dernonextsusy} as well.

\section{Higher derivative corrections to the gravitational index}\label{sec:higherdersol}

In this section we present the rotating black hole solution including higher-derivative corrections in the~$\CN=2$ supergravity coming from string theory. 
The higher-derivative corrections in~$\CN=2$ supergravity are of two types: chiral superspace integrals 
of the holomorphic prepotential function, and full superspace integral, sometimes called F-terms and D-terms, respectively, 
in the~$\CN=2$ context. We only consider the first type of terms, which are 
summarized by corrections to the tree-level holomorphic prepotential function~$\CG(\CY^I)$ 
discussed in the previous section. We then make comments on D-terms and the controllability of the solution.

The treatment begins by introducing another chiral field~$\wh A$ of weight $2$, whose top component is proportional 
to the Weyl-squared tensor in four-dimensions.   
The prepotential is extended to a function~$F(X^I,\wh A)$ and, in addition to the first derivatives~$F_I=\partial F/\partial X^I$, we have $F_{\wh A}=\partial F/\partial \wh A$. 
The particular form of the prepotential 
that arises in Calabi-Yau three-fold compactifications of Type II string theory is 
\be \label{fullF}
F(X^I,\wh A) \= -\frac{1}{6}\frac{C_{IJK} X^I X^J X^K}{X^0} -c\frac{\wh A}{64} \frac{X^1}{X^0} + \dots \,.
\ee
Here~$\wh A$ effectively counts the loops in the topological string expansion. 
As~$\wh A$ contains four derivatives, we can also think of the series as an expansion in the number of derivatives.

The cubic or tree-level term was discussed in the previous section, and we have shown the one-loop term. 
Here~$c=c_2/24$ where $c_2$ is the second Chern class of the Calabi-Yau. 
In addition there are typically an infinite series of terms
given by the higher loops of the topological string expansion. 
In Type II theory on~$K3 \times T^2$, the term~$X^1/X^0$ above is completed by worldsheet instanton
effects to a modular function of~$X^1/X^0$. Importantly, the expansion terminates 
at this order due to the fermion zero modes on~$T^2$.

In our case of interest here, i.e.~the two-charge black hole, the 
two-derivative result is singular, so that in the full solutions the correction terms are as important as the two-derivative  
term. This means that, a priori, one does not have a controllable expansion. 
Nevertheless it is useful to see explicitly that  including higher-derivative terms can lead to a smooth solution, as we show below.

A proof that D-terms vanish on configurations that preserve supersymmetry 
would justify the approach where we only keep F-terms.\footnote{As opposed to previous work \cite{Sen:2004dp,Hubeny:2004ji,Dabholkar:2004dq}, we do not expect the generic expansion to be controllable for the thermal partition 
function which does not include an insertion of $(-1)^{\sf F}$.} 
In some situations, like the attractor black hole entropy, one can indeed show that D-terms do not affect the quantity of 
interest~\cite{Butter:2014iwa, Murthy:2013xpa}, and therefore one obtains an exact formula~\cite{Dabholkar:2010uh, Dabholkar:2011ec, Dabholkar:2014ema, Iliesiu:2022kny}. 
In our present context, as we see below, the entropy arising from the 
prepotential~\eqref{fullF} agrees with the microscopic string theory index including the numerical coefficient, pointing to a non-renormalization result.

The analysis of supersymmetry equations and solutions of the higher-derivative theory was performed 
in the impressive set of papers~\cite{LopesCardoso:1999za, LopesCardoso:2000qm, LopesCardoso:2000fp}. 
Just as we define the rescaled fields~$\bigl(\CY^I, \CG_I \bigr)$ in~\eqref{defYG}, we also define a rescaled 
field~$\Upsilon= e^{\CK(X, \overline{X})} \, \overline{Z}^2 \wh A$. The higher-derivative equations are, generally speaking, more complicated than the two-derivative theory.  
The exception is the equations governing the scalar fields, which are still given by the generalized stabilization equations~\eqref{genstab}, although they now depend on $\Upsilon$.

\subsection*{New attractor with higher-derivative corrections}

The solution is specified by a vector of harmonic functions $H=(\wt H^I , H_I)$. The generalized stabilization equations are now 
\be \label{genstab222}
\CY^I - \overline{\CY}^I \= \ii \wt H^I \,, \qquad 
\CG_I(\CY,\Upsilon) -  \overline{\CG}_I(\CY,\Upsilon)  \= \ii H_I  \,,
\ee
and they determine $\CY^I$ in terms of $H$ and $\Upsilon$. The metric is of the IWP form as in the two-derivative theory 
\be
\d s^2 \= e^{2U} ( \d t_{\rm E}+\i \omega_m \d x^m)^2 + e^{-2U} \d x^m \d x^m \,.
\ee 
The functions~$U$ and~$\omega$ are now governed by 
\be \label{UEq}
\begin{split}
e^{-2U} &\= \ii \bigl( \CY^I\overline{\CG}_I  - \CG_I \overline{\CY}^I \bigr) + 128 \ii \, e^{U} \nabla^p \bigl(  (\nabla_p e^{-U})(\CG_\Upsilon - \overline{\CG}_\Upsilon) \bigr) \\
&\quad -32 \ii \, e^{4U} (R(\omega)_p)^2 (\CG_\Upsilon - \overline{\CG}_\Upsilon)-64 e^{2U} R(\omega)_p \nabla^p (\CG_\Upsilon + \overline{\CG}_\Upsilon) \,.
\end{split}
\ee
\be \label{OmegaEq}
\begin{split}
-i R(\omega)_p &\=H_I\overset{\leftrightarrow}{\nabla}_p  H^I -128 \i \, \nabla^q \left[ \nabla_{[p}(e^{2U} R(\omega)_{q]} (G_\Upsilon - \overline{G}_\Upsilon))\right] \\
&\quad -128 \, \nabla^q \bigl( 2 \nabla_{[p} U \, \nabla_{q]} (G_\Upsilon + \overline{G}_\Upsilon) \bigr) \,.
\end{split}
\ee
In the two-derivative theory $\CG_\Upsilon=0$ and the above equations 
reduce to the simpler equations~\eqref{Uomgeneral}. Finally, the value of $\Upsilon$ is 
\be\label{UpsilonEq}
\Upsilon \= -64 \Bigl(\nabla_p U - \frac{\ii}{2} e^{2U} R(\omega)_p \Bigr)^2 \,. 
\ee
Here we follow the notation of~\cite{LopesCardoso:1999za} 
with~$\nabla_a$ the derivative in the 
three-dimensional flat base-space, and $R(\omega)^p = \varepsilon^{pqs} \nabla_q \omega_s$. For the non-extremal rotating saddle we set, as in Section~\ref{sec:4dsugra}
\be \label{HNSsplit222}
\Hv(\xvec) \= h + \frac{\gamma_N}{|\xvec-\xvec_N|} + \frac{{\gamma}_S}{|\xvec-{\xvec}_S|}\,, \qquad \gamma_N + \gamma_S \= \Gamma \,.
\ee
Therefore part of the problem is to determine $\gamma_{N}$, $\gamma_S$ and $|\xvec_N - \xvec_S|$ using smoothness, just like we did at the two derivative level \cite{Boruch:2023gfn}.

\bigskip

As an exercise we can study the behavior of all the fields in 
an extremal attractor background as in the discussion near~\eqref{Hextr}, as~$r \to 0$,
\be
\text{Extremal:}\qquad R(\omega) \= 0 \,, \qquad \CY^I \; \sim \; \frac{Y_*}{r} + \dots\,, \qquad 
e^{-2U} \; \sim \; \frac{a}{r^2} + \dots\,, \qquad \Upsilon \; \sim \; \frac{\Upsilon_*}{r^2} + \dots\,.
\ee
The fact that $U \sim \log r$ together with \eqref{UpsilonEq} implies that~$\Upsilon_*=-64$. The scalars $Y_*^I$ are determined by $Y_*^I-\bar{Y}_*^I = \i P^I$ and $G_{I*}-\bar{G}_{I*}= \i Q_I$.
The scalar fields therefore behave as in the two-derivative theory. The scaling 
behavior of the other fields can be read off from their conformal weights---this can be explained either by using a conformal compensator, or by thinking of the other fields as functions of the scalar fields.
The only thing to be careful of is that coefficients can be zero, as happens for~$R(\omega)$ because of the spherical symmetry. 
The value~$\Upsilon_*=-64$ is precisely the one required for the Wald entropy of the 
extremal black hole to equal the logarithm of the growth of microscopic states of the fundamental heterotic string~\cite{Dabholkar:2004yr}.

\bigskip

Now we repeat the analysis for the rotating non-extremal solution, and determine the value of $\gamma_{N/S}$. We focus on the behavior of the solution near the north pole, since the behavior at the south pole is related by complex conjugation. Recall that the idea in the rotating non-extremal solution is to demand that the metric factor $e^{-2U}$ has only a single pole, leading to a finite temperature horizon. 
Accordingly, we assume the following behavior for the non-extremal solution, as~$\rho = |\xvec-\xvec_N| \to 0$, 
\be
\CY^I \; \sim \; \frac{Y_N}{\rho} + \dots\,, \qquad 
e^{-2U} \; \sim \; \frac{a_1}{\rho} + \dots\,, \qquad 
R(\omega)_p \= \frac{a_2}{\rho^2}\, \wh n + \dots \,, \qquad 
\Upsilon \; \sim \; \frac{\Upsilon_*}{\rho^2} + \dots\,,
\ee
where~$\wh n = (\xvec-\xvec_N)/|\xvec-\xvec_N|$. 
Now we analyze the equations~\eqref{UEq}--\eqref{UpsilonEq}. The upshot is that the solution to the extremal attractor~$(Y_*^I, G_{I*})$
together with~$\Upsilon_*=-64$ determines~$\gamma_{N}$.
Together with the smoothness condition, this completely determines the sources~$H$. 
Now the full solution can be determined in terms of the boundary data ($\Gamma$ and~$\beta$)\footnote{We assume that, given $H$, 
a solution to the partial differential equations \eqref{UEq} and \eqref{OmegaEq} exists and is unique. This has been verified in the low temperature limit numerically.}.

We begin with~\eqref{OmegaEq}. 
%From~\eqref{Gupssol} we see that~$\CG_\Upsilon$ is purely imaginary, so 
%that the last term in~\eqref{OmegaEq} vanishes. 
$\CG_\Upsilon$ has Weyl weight~0 and so $\CG_\Upsilon \sim \text{O}(1)$ as~$\rho \to 0$. 
The leading behavior of the second term on the right hand side is therefore proportional to~$\nabla_{[p}(e^{2U} R(\omega)_{q]})$. Now, $e^{2U} R(\omega)_q \sim \frac{a_2}{\rho} \wh n_q + \dots$, 
and therefore the anti-symmetry of the action of the derivatives annihilates this term 
so as to give no contribution at~$\text{O}(1/\rho^2)$. A similar analysis applies to the last term in \eqref{OmegaEq}.
This shows that the value of~$R(\omega)$ at the pole in the higher-derivative theory is 
the same as its value given in~\eqref{Romegaval} in the two-derivative 
theory. We conclude that~$a_2 = i \langle \gamma_N, H^{(0)}\rangle$. 
In particular, this means that the condition of there being no Dirac-string singularity 
implies the same relation~\eqref{gensmoothcond} as in the two-derivative theory, which we repeat here for convenience 
\beq
\frac{\i \langle \gamma_N, \gamma_S \rangle}{|\xvec_N - \xvec_S|} + \i \langle \gamma_N, h\rangle \= \frac{\beta}{4\pi} \,.
\eeq

Now we move to~\eqref{UEq}. Let us start by considering the first term on the right hand side. 
This is the value of $e^{-2U}$ in the two-derivative theory.
We consider a solution that cancels the double pole coming from this term by imposing $\bar{Y}_N^I=\bar{G}_{IN}=0$ (and $Y^I_S=G_{IS}=0$ for the south pole). 
The stabilization equation, together with the old attractor, would then imply
\beq\label{eq:gammaNhdc}
\i \gamma_N \= (Y_*^I, G_{I*}).
\eeq
This solution is not yet complete, as 
we have not yet determined $\Upsilon_*$ which appears implicitly in $G_I$. We will come back to this point shortly. 
First, we explain how the other terms in~\eqref{UEq} lead to a vanishing double pole. 
As~$\rho \to 0$, calling the constant~$\CG_\Upsilon = C_1$, the leading behavior of the  second and third terms is governed by 
\be
C_1 
\bigl(128 \ii \, e^{U} \nabla^2 e^{-U} -32 \ii \, e^{4U} (R(\omega)_p)^2
\bigr) \; \sim \; 
-32 \ii \, C_1 \Bigl( 1+\frac{a_2^2}{a_1^2} \Bigr) \frac{1}{\rho^2} + \dots \,.
\ee
We see that in order to have a vanishing double pole we must have~$a_2 = \pm i a_1$.
In the two-derivative theory we have, from~\eqref{e2UNP} and~\eqref{YNvals}, the 
value $a_1=-\langle \gamma_N, H^{(0)}\rangle$ so that~$a_2 = i a_1$. 
We choose this solution also in the higher-derivative theory. 
This implies a condition of vanishing of other~$\text{O}(1/\rho)$ terms of the right-hand side of~\eqref{UEq} involving higher derivative terms. From the equation~\eqref{UpsilonEq}, we deduce that   
\be 
\Upsilon \; \sim \; -\frac{64}{\rho^2} 
\Bigl(\frac12 - \frac{\ii}2 \frac{a_2}{a_1}\Bigr)\; \= \; -\frac{64}{\rho^2}\,. 
\ee
This completes the derivation since now we determined $\Upsilon_*=-64$. As in the two-derivative solution, we see that the holomorphic (anti-holomorphic) scalar fields take the attractor values at the north (south) pole. Furthermore they determine the dipole charges $\gamma_N$ ($\gamma_S$) through the relation \eqref{eq:gammaNhdc}. 

So far we described the metric and scalar profiles. The gauge fields are also determined by the harmonic sources through 
the equations $F^I_{0p} = - \nabla_p [ e^{2U} (\CY^I + \bar{\CY}^I)]$ and $G_{0p}^I = - \nabla_p [ e^{2U} (\CG_I + \bar{\CG}_I)]$ 
where $F^I_{\mu\nu}$ and $G^I_{\mu\nu}$ determine the field strength and its dual. These equations are the same as in the 
two-derivative theory.

To conclude the characterization of the solution, we propose a formula for the on-shell action of such configurations, 
including boundary terms that remove the zero-point energy~$\beta M_{\rm BPS}$, to be
\beq\label{eq:ansatzactionhdt}
-I_{\rm on-shell} \= \pi (q_I Y^I_N + 2\i \CG(Y_N)) + \pi (q_I \overline{Y}_S^I -2\i \overline{\CG(Y_S)} ).
\eeq
We have two reasons that justify this expression. First, it reduces to the two-derivative action~\eqref{IGGbar} when $\CG$ is independent of $\Upsilon$, for any temperature. (Note that when $\CG$ depends on $\Upsilon$ \eqref{IGGbar2der} is different from \eqref{IGGbar}.) 
Second, one can prove this formula in the presence of arbitrary higher-derivative corrections, under the assumption that the on-shell action is temperature independent. 
The action in the zero-temperature limit was explicitly evaluated in \cite{Dabholkar:2010uh} and matches \eqref{eq:ansatzactionhdt} (as well as the Wald entropy derived in \cite{LopesCardoso:1999za}). 

The reason we have not attempted an explicit evaluation of the action is that it requires solving the differential equations \eqref{UEq} and \eqref{OmegaEq}. Nevertheless, there might be a more elegant way to obtain \eqref{eq:ansatzactionhdt}. 
The split form of the action  in~\eqref{eq:ansatzactionhdt} is similar to observations of the factorization of the partition function in~\cite{BenettiGenolini:2019jdz, Hosseini:2019iad}.  
An efficient way to prove this could be to apply the techniques of~\cite{BenettiGenolini:2023kxp}, we leave this for future work.

\subsection*{The corrected two-charge solution}
We now come back to the specific theory relevant to the two-charge black hole in heterotic string theory, 
with only $X^1,X^2=X^3$, with 
% non-zero. The prepotential is therefore 
prepotential 
\beq
F(X^I,\wh A) \= -\frac{X^1 (X^2)^2}{X^0} - c \frac{\wh A}{64} \frac{X^1}{X^0} \,.
\eeq
For the heterotic string $c=1$ but we find it useful to keep it general to keep track of the effect of higher derivative corrections. The non-zero monopole charges in the solution are taken to be~$Q_0, P^1$ and recall that $Q_0<0$ and $P^1>0$. 
According to the discussion in Sections~\ref{sec:newattr},~\ref{sec:2charge},
we choose sources $(\wt H^0, \wt H^1, \wt H^2, H_0, H_1, H_2)$.
Let us begin by solving the new attractor equations which determine $\gamma_{N/S}$. The 
starting point is the extremal attractor solution
\beq
Y_*^I \= \left(\sqrt{\frac{cP^1}{|Q_0|}},\frac{\i P^1}{2},0\right)\,, \qquad 
G_{I*} \=\left(\frac{\i Q_0}{2}, -\sqrt{\frac{cP^1}{|Q_0|}},0\right)\,, \qquad 
\Upsilon_* \= -64 \,,
\eeq
which is equal to $Y_N$ at the north pole while $\bar{Y}_N=0$. The divergence in the dilaton (near the north and south pole) at the horizon is now regulated since $Y_*^0$ is finite.
The fact that now $G_{2*}=0$ 
and $G_{1*}\neq 0$ might seem like a major modification compared to the two-derivative level discussion. 
To see how this comes about one can turn a small charge $P^2$ (or, equivalently, go slightly away from the pole, 
as explained in Section \ref{sec:2charge}). 
For example, this produces a term $G_{2*} \sim \sqrt{P^1|Q_0|} \frac{P^2}{4c + (P^2)^2}$. 
We see explicitly that the limits $c\to0$ and $P^2\to0$ do not commute and this is the source of discrepancy. 
This is clearly due to the singular nature of the two-derivative solution and would not happen for a big black hole.

According to the new attractor solution~\eqref{eq:gammaNhdc}
the charge vectors associated to the north and south pole contributions are
\beq\label{gammaNhdc}
\gamma_N \= \left(\i\sqrt{\frac{cP^1}{|Q_0|}}, \frac{P^1}{2},0,\frac{Q_0}{2}, - \i\sqrt{\frac{cP^1}{|Q_0|}},0\right)
\eeq
and $\gamma_S = (\gamma_N)^*$. Notice again that this is not a small correction to the 
two-derivative solution for $\gamma_N$!  One can check that now we have  
\beq
\i\langle \gamma_N, \gamma_S \rangle \= 2 \sqrt{c P^1 |Q_0|} \,.
\eeq

The final ingredient to determine the sources \eqref{HNSsplit222} and therefore the full solution is to choose $h$. 
At infinity we have $\Upsilon \to 0$ and therefore we can repeat the same analysis as in the two-derivative theory to establish the values $h^I= \wt H^I|_{\infty}$ and $h_I= H_I |_{\infty}$ 
that allow us to compare with the solution in Sections~\ref{sec:SBHthermo} and~\ref{sec:2dernonextsusy}. 
The answer found in Section~\ref{sec:2charge} is $h_I=(-1,0,0)$ and $h^I=(0,1,\frac12)$. 
The solution to the generalized stabilization equations~\eqref{genstab} with prepotential~\eqref{fullF} becomes then an algebraic equation that relates $(\CY^I, \CG_I)$ to $(\wt H^I, H_I)$ and $\Upsilon$, 
which can be solved easily\footnote{It is important to emphasize that the stabilization equations are not enough 
to determine~$\CY$ in terms of~$H$ in the higher-derivative theory. The reason is that they involve $\Upsilon$, which depends on $U$ 
and $\omega$. Therefore the stabilization equations are not decoupled from \eqref{UEq} and \eqref{OmegaEq}. }.

\bigskip

The relation \eqref{gensmoothcond}  fixes 
the parameter~$2{\sf a}=|\xvec_N-\xvec_S|$ in terms of~$\beta$, as  anticipated earlier in Section \ref{sec:2dernonextsusy}. 
Since, in the language of that section, $\i\langle \gamma_N , \gamma_S \rangle = 2 \sqrt{c P^1|Q_0|}= 2\sqrt{c} m_0$ and $\i\langle \gamma_N, h \rangle = \frac{\sqrt{c}(P^1-Q_0)}{\sqrt{P^1 |Q_0|}}=2 \sqrt{c} \cosh \upalpha $, 
the  relation~\eqref{gensmoothcond} gives
\beq \label{betaarel}
\frac{2\sqrt{c} m_0}{2 {\sf a}} + 2 \sqrt{c} \cosh\upalpha \= \frac{\beta}{4\pi} \,, \qquad 
\Rightarrow~~{\sf a} \=\frac{4\pi m_0}{\frac{\beta}{\sqrt{c}}-8 \pi \cosh \upalpha} \, .
\eeq
This relation resolves the problem appearing in the two-derivative expression~\eqref{eq:tempsmallbh}, 
namely that the temperature was apparently fixed in terms of the charges and asymptotic moduli.
Now we see that the temperature and the moduli are independent quantities. 
The relation~\eqref{betaarel} shows the solution is only valid as long as the denominator is positive.
Curiously, the denominator changes sign at a finite value of~$\beta$. (The same 
phenomenon is found even for the big black hole where $|\xvec_N-\xvec_S|$ becomes negative above a charge-dependent temperature. 
Explaining this remains an open question.)

In summary, we find a solution that is smooth  with the dilaton finite everywhere, 
all the parameters of the solution are determined in terms of boundary data which are 
independent values of the temperature and moduli.

\bigskip

To conclude this section we  evaluate the on-shell action of the solution we constructed, using \eqref{eq:ansatzactionhdt}. Since $Z_{\rm index} \sim e^{-I_{\rm on-shell}}$ we find that it makes the following contribution to the gravitational index
\bea
\log Z_{\rm index} &\=& \pi (q_I Y^I_N + 2\i \CG(Y_N)) + \pi (q_I \overline{Y}_S^I -2\i \overline{\CG(Y_S)} ) \,,\\
&\=& 4 \pi \sqrt{ c P^1 |Q_0|} \,.
\ea
This is now finite. To compare with the heterotic string action we recall that $c=1$. Therefore we find a 
gravitational index given, in terms of the charges $n=-Q_0$ and $w=P^1$, by 
\beq
\log Z_{\rm index} \= - I_\text{on-shell} \= 4 \pi \sqrt{nw} \,.
\eeq
(In terms of the parameter $m_0$ of Section \ref{sec:2dernonextsusy} this is $ 4\pi m_0$.)
This result precisely matches the microscopic evaluation of the index counting BPS fundamental heterotic strings, as reviewed in the introduction. 

Note that the same phenomenon is known to occur if one assumes an attractor AdS$_2$ geometry~\cite{Dabholkar:2004yr}. 
What is remarkable here is that one obtains the same result even though the rotating geometry for the index 
preserves much less symmetry, and is compatible with the discussion in Section \ref{sec:SBHthermo} on the partition function.

\section{Can the low-temperature physics be nearly conformal?}\label{sec:canitbe}

In Section~\ref{sec:SBHthermo} we studied two-charge black holes in string theory. At low temperatures the black hole becomes small, i.e.~it has a horizon area that vanishes at the two-derivative level.
Following \cite{Chen:2021dsw}, we explained how a transition takes place to a winding condensate without an event
horizon, before we reach the small black hole,  
and ultimately the ground state is described by a gas of strings and not a black hole. 

Is the behavior described above universal? Could there be a string theory solution that does not 
transition to a gas of strings or other stringy objects without a horizon, such that the ground 
state is really described by a small black hole?
In this section we use effective theory arguments to comment on the possibility of a string-size near-extremal black hole. Our arguments also apply for big black holes with 16 or more supercharges. 

By a small black hole we mean a gravitational solution with a finite (string scale) area that exists at very low temperatures, which develops an ${\rm AdS}_2$ throat. 
This assumption can be justified in some cases where 
there is a smooth limit to the two-derivative theory~\cite{Kunduri:2007vf,Figueras:2008qh}. 
% We assume that there is an~$SU(2)$ rotational symmetry in four dimensions,
% consistent with 
% of the string compactified to four dimensions 
We also assume that the index is a function that grows exponentially with the charges. 
The latter requirement is satisfied by most examples in string theory (an exception being e.g.~the D0 brane black hole). 
Finally, we assume that the extremal limit of the small black hole preserves at least~16 supercharges.

The near-extremal AdS$_2$ phase of such black holes implies the presence of a nearly-conformal symmetry of the quantum system describing the black hole microstates. 
Its low energy description is given by the Schwarzian theory.
Since we assume the system preserves a certain number of supercharges, we have a supersymmetric generalization of the  Schwarzian theory.

It is possible to label the different supersymmetric extensions of the Schwarzian theory by the 
superconformal group of symmetries broken by the 
near-AdS$_2$ solution. 
For example the bosonic Schwarzian theory would be labeled by ${\rm PSL}(2,\mathbb{R})$, the one describing the near-BPS limit of the 1/16-BPS black holes in ${\rm AdS}_5$ 
would be labeled by $\SU(1,1|1)$ and the one describing the near-BPS limit of large black holes in flat space by ${\rm PSU}(1,1|2)$.
Therefore we can address this problem by analyzing the possible superconformal groups. 
To perform our analysis we use that the supersymmetric Schwarzian action acts as a ${\rm U}(1)$ generator 
in the space of super-reparametrization, such that we can apply the localization principle of~\cite{Stanford:2017thb}.
We assume that the R-symmetry of the superconformal group is realized in a geometric manner in spacetime.\footnote{These assumptions do not rule out the possibility of a sigma model description. 
See~\cite{Giveon:2006pr,Dabholkar:2007gp, Johnson:2007du,Lapan:2007jx,Kraus:2007vu} for a 
related discussion on 
the~AdS$_3$ near-horizon configurations of a fundamental string.}  
We now show that no Schwarzian theory exists with these properties.

Given the choice of a superconformal group we can derive the partition function of such a theory including quantum corrections. 
These corrections are one-loop exact and have the following form. 
Denote by~$\vec{\mu}$ the vector of chemical potentials that couple to the $r={\rm rank}(G)$ 
Cartan generators of the $R$-symmetry group $G$, 
with the periodicity~$\mu \sim \mu+1$. The partition function then takes the form
\beq
Z \= \sum_{\vec{n} \in \mathbb{Z}^r} \beta^{\frac{N_F-N_B}{2}} \, Z_{\rm one-loop}(\vec{\mu}+\vec{n}) \, e^{S_0 + \frac{1}{\beta} I(\vec{\mu}+\vec{n})} \,,
\eeq
where $Z_{\rm one-loop}(\vec{\mu}+\vec{n})$, and~$I(\vec{\mu}+\vec{n})$, are functions of chemical potentials 
only arising from the one loop determinant, and on-shell action, respectively. The sum over integers $\vec{n}$ is a sum over saddles related by large gauge transformations. 
$S_0$ is the extremal entropy while $N_F$ and $N_B$ correspond to the number of fermionic generator and bosonic generators in the superconformal group. 
The overall prefactor of $\beta^{\frac{N_F-N_B}{2}}$ in the one-loop determinant comes from subtracting a contribution from zero-modes. 

When $N_F\geq N_B$ the theory can predict the presence of a large number of ground states of order $e^{S_0}$. 
%For example $Z \sim e^{S_0} + O(e^{-\beta})$. 
For example for $\SU(1,1|1)$, $N_B-N_F=0$ so that the one-loop determinant is temperature independent 
and in the zero-temperature limit the action becomes temperature independent, 
resulting in $Z(\beta \to\infty) \sim e^{S_0}+ O(e^{-\beta})$. 
For ${\rm PSU}(1,1|2)$, instead, $N_F=8>N_B=6$ and the one-loop determinant diverges as $\beta \to \infty$. In this case the sum over saddles is important in order to regulate this divergence and once again we find $Z(\beta \to \infty) \sim e^{S_0}$.

When $N_B>N_F$ the one-loop determinant vanishes in the $\beta \to\infty$ limit and its unavoidable to obtain $Z(\beta \to \infty) =0$. 
This implies that the quantum corrections due to the Schwarzian mode would destroy the classical degeneracy of ground states. In other words $\log Z \sim S_0 - \frac{N_F -N_B}{2} \log \beta \to -\infty$ in the large $\beta$ limit. 
Since the index gives a lower bound on the number of ground states, 
we can use this criterion to rule out the existence of a small black hole horizon with an ${\rm AdS}_2$ factor.  
In other words, there is no mass-gap and therefore no decoupled~AdS$_2$ geometry in such theories.

By the arguments in the previous paragraph we reduced the problem to an analysis of the number of 
bosonic vs fermionic generators of superconformal groups~\cite{Nahm:1977tg}. 
The list of such groups can be found e.g. in 
Table~1 of the review~\cite{Britto-Pacumio:1999dnb}, see~\cite{Frappat:1996pb} for an extensive list.

Let us begin with superconformal groups with more than $16$ supercharges. There are three possible families. The first family are $\SU(1,1| n)$ with $n>4$, which has a $\SU(n) \times {\rm U}(1)$ $R$-symmetry. 
This group has $N_B=n^2+3$ and $N_F = 4n$. For all $n>1$ one has $N_B >N_F$ and therefore the ground states do not survive quantum corrections. 
The same is true for ${\rm OSp}(n|2)$ for $n>8$ since $N_B = \frac{1}{2} (n^2-n) + 3$ and $N_F = 2n$ and for ${\rm OSp}(4^*|2n)$ with $n>2$ which has $N_B = 2n^2 + n +6$ and $N_F = 8n$. Therefore, if an index predicts a large ground state degeneracy, those states cannot describe small black holes preserving more than 16 supercharges. 

Next, we consider the case of systems with 16 supercharges. In this case there are four superconformal groups, which are 
\begin{equation}
\hspace{-0cm}\begin{tabular}{S|SS} \toprule
    {\footnotesize  Superconformal Group} & {\footnotesize  $N_B$/$N_F$} & {\footnotesize  $R$-symmetry}   \\ \midrule\midrule
  {\footnotesize  ${\rm OSp}(4^*|4)$} & {\footnotesize  $16/16$} & {\footnotesize  $\SU(2)\times {\rm Spin}(5)$}  \\ %\midrule
  {\footnotesize  $\SU(1,1|4)$} & {\footnotesize  $19/16$}  & {\footnotesize  $\SU(4) \times {\rm U}(1)$}  \\  %\midrule
  
    {\footnotesize  $F(4)$}  & {\footnotesize  $24/16$}  & {\footnotesize  $\SO(7)$}   \\
    {\footnotesize  ${\rm OSp}(8|2)$}  & {\footnotesize  $31/16$}  & {\footnotesize  $\SO(8)$}   \\\bottomrule
\end{tabular}
\end{equation}
The last three of these groups have $N_B>N_F$ and therefore do not have a ground state degeneracy. 
The case ${\rm OSp}(4^*|4)$ is special since $N_B=N_F$. 
Here, the~${\rm Spin}(5)$ R-symmetry does not act as a geometric rotation of any 
subspace of~$T^6$.\footnote{It is possible 
that the supercharges 
transform as a spinor of the tangent space~${\rm Spin}(5)$. This possibility was discussed in related~AdS$_3$ near-horizon configurations in~\cite{Dabholkar:2007gp}.}

Thus, there is no superconformal group, and therefore no Schwarzian 
theory, which could describe the low energy phase of such small black holes. This argument can be 
extended to theories with 14 supercharges. Below 14 there are multiple possibilities of nearly-conformal 
theories with a large number of ground states such as $\SU(1,1|1)$ or ${\rm PSU}(1,1|2)$, and other 
superconformal groups which haven't found an application yet. Another theory that deserves further study along these lines is~${\rm D}(2,1;\alpha)$.

\section{Discussion \label{sec:discussion}}

We would like to conclude with some comments and future directions.

We have evaluated the gravitational path integral that corresponds to the index of the two-charge states in heterotic string theory. 
This calculation involved constructing a solution that combines two-derivative with four-derivative interactions. 
An obvious future direction is to make this construction more precise. 
The advantage of the present paper compared with previous attempts is that we turned this question into a concrete program, 
starting with proving that D-terms vanish due to supersymmetry. Besides this, it would be necesasry 
to prove that the action coming from F-terms is temperature independent and reproduces \eqref{eq:ansatzactionhdt} for any prepotential.

Another natural direction is to compute quantum corrections around the black hole solution evaluating the index. We hope that one can use the recent techniques in~\cite{H:2023qko} and~\cite{Anupam:2023yns} to reproduce the logarithmic correction to the index in the large $N=\sqrt{nw}$ limit. An even more ambitious goal can be to use supergravity localization to reproduce the exact generating function $1/\eta^{24}(\tau)$. 

Finally, it is interesting to consider whether this approach can be extended to black holes with one charge. An example is a 10 dimensional type IIA string theory black hole with D0-charge $N$. This leads to a small black hole in the extremal limit and its near-extremal excitations are dual to the BFSS matrix model. In this case even a microscopic calculation of the index is hard and has only been completed for the $N=2$ theory \cite{Sethi:1997pa}, although there is a simple M-theory prediction for it for all $N$. To attack this problem using the gravitational path integral, one can easily find a rotating D0-brane black hole starting with the uncharged Myers-Perry solution in 10 dimensions $(t,x^1,\ldots,x^9)$, uplifting to 11 dimensions by adding an extra circle $x^{10}$, performing a boost in the $t-x^{10}$ coordinates, and finally dimensionally reducing back to 10 dimensions. Imposing the boundary conditions relevant for the index one can find, at the two-derivative level, a supersymmetric rotating non-extremal black hole with a macroscopic area and zero on-shell action, similar to its two-charge counterpart of Section~\ref{sec:2dernonextsusy} \cite{gjtwip}. The solution is singular at the poles of the horizon and the string coupling becomes large at those locations. An evaluation of the index of the BFSS model using the gravitational path integral for type IIA seems to require a new idea and cannot be computed by a simple extension of the techniques used here.

\bigskip

\section*{Acknowledgements}

It is a pleasure to thank Pietro Benetti-Genolini, Jan Boruch, Matt Heydeman, Luca Iliesiu, Gabriel Lopes Cardoso, 
Juan Maldacena, Thomas Mohaupt, Ashoke Sen, and Edward Witten
for interesting and useful discussions. Y.C. acknowledges the KITP program ``What is String Theory? Weaving Perspectives Together", during which this work is completed.
S.M.~acknowledges the support of the J.~Robert Oppenheimer 
Visiting Professorship at the Institute for Advanced Study, Princeton and 
the STFC grants ST/T000759/1,  ST/X000753/1.
The work of GJT is supported by the University of Washington and the DOE award DE-SC0024363.

\bigskip

\appendix

\section{Solution generation for rotating two-charge systems}\label{sec:solngene}

In this appendix, we discuss how one can use the solution generating procedure to work out the properties of the rotating two-charge system from the uncharged (but rotating) solution. We will derive the thermodynamic quantities of the charged solution with the information of the thermodynamic quantities of the uncharged solution as well as the transformation at infinity. The derivation applies to general classical solutions in string theory, including black hole solutions as well as potential rotating Horowitz-Polchinski solutions that have not yet been constructed. One application of our derivation, which we discuss in appendix. \ref{sec:alphacorr}, is that one can work out the $\alpha'$ corrections to the rotating two-charge black hole in \cite{Sen:1994eb}, without needing to find the corrected geometry explicitly. The discussion here mostly follows the derivation in \cite{Chen:2021dsw}, with the extra ingredient that the seed solution can carry rotation. 

The starting point is a seed solution in $D$ dimensions which has an asymptotic time circle with length $\tilde{\beta} = 2\pi \tilde{R}$. For the purpose of applying to the discussion in the main text, we have $D=4$, though the discussion is more general. We will assume the existence of a $\U(1)$ isometry, time translation, everywhere in the solution. For simplicity, we consider the transformations that generate charges along one extra circle, and the relevant symmetry group is $O(2,2)$ which acts on fields in $d=D-1$ dimensional space once we dimensional reduce on both the time circle and the extra circle. This is a subgroup of the $O(7,23)$ transformation of heterotic string theory compactified on $T^6$ we considered in the main text, but it also applies to the Bosonic/Type II theories compactified on a torus. We parametrize the transformation by the parameters $(\upalpha,\upbeta)$.

The main equation, which we refer to \cite{Chen:2021dsw} for justification, is the invariance of the action under the solution generating transformation.
\begin{equation}\label{eqac}
	\log Z(R,\Omega,r,\mu_n,\mu_w,\Phi_D) = \log \tilde{Z} (\tilde{R},\tilde{\Omega}, \tilde{\Phi}_D).
\end{equation}
Let's unpack this equality by explaining the notations. All the tilded quantities refer to the seed solution, which has an angular potential $\tilde{\Omega}$ other than the inverse temperature (over $2\pi$) $\tilde{R}$ and asymptotic value of the dilaton $\tilde{\Phi}_D$. The quantities without tildes are those for the generated solution, which also has $r$ being the size of the extra circle, and $\mu_n, \mu_w$ being the chemical potentials for the momentum and winding charges $Q_n$ and $Q_w$. We will fix $r=1$ in the following discussion.\footnote{This is only valid since we are using $r$ to denote the asymptotic size of the circle. Inside the spacetime it will be varying.} We can factor out the dependence of the dilaton and define the primed quantities\footnote{For the dilaton, here we follow the convention used in the main text, see (\ref{action}). }
\begin{equation}
	e^{-\Phi_D} \log Z' (R,\Omega,\mu_n,\mu_w) \equiv \log Z(R,\Omega,\mu_n,\mu_w,\phi_D), \quad e^{-\tilde{\Phi}_D}\log \tilde{Z}' (\tilde{R},\tilde{\Omega}) \equiv \log \tilde{Z} (\tilde{R},\tilde{\Omega}, \tilde{\Phi}_D)
\end{equation} 
Using the invariance of the $d=D-1$ dimensional dilaton, the asymptotic value of the dilaton transforms as
\begin{equation}
	e^{-\tilde{\Phi}_D} \tilde{R} = e^{-\Phi_D} R
\end{equation}
and therefore from (\ref{eqac}) we have
\begin{equation}
	\log Z' = \frac{R}{\tilde{R}} \log \tilde{Z}'.
\end{equation}
For the generated solution, we have the general statistical expression for the action
\begin{equation}
	\log Z = e^{-\Phi_D} \log Z' = S - 2\pi R M + 2\pi R \Omega J + 2\pi R \mu_n Q_n + 2\pi R \mu_w Q_w
\end{equation}
where various charges and entropy can be computed as
\begin{equation}\label{charges}
\begin{aligned}
	& 2\pi R Q_n = e^{-\Phi_D} \partial_{\mu_n} \log Z',\quad 2\pi R Q_w = e^{-\Phi_D} \partial_{\mu_w} \log Z', \quad 2\pi R J = e^{-\Phi_D} \partial_{\Omega} \log Z', \\
	& 2\pi M = e^{-\Phi_D} \left( - \partial_R + \frac{\mu_n}{R} \partial_{\mu_n} + \frac{\mu_w}{R} \partial_{\mu_w} + \frac{\Omega}{R} \partial_{\Omega}\right)\log Z', \quad S = e^{-\Phi_D} (1- R\partial_R) \log Z'.
\end{aligned}
\end{equation}
On the other hand, for the seed solution, we have
\begin{equation}\label{primerelations}
\begin{aligned}
&\log \tilde{Z}' =\tilde{S}' - 2\pi \tilde{R} \tilde{M}' + 2\pi \tilde{R} \tilde{\Omega} \tilde{J}', \\
	& 2\pi \tilde{R} \tilde{J}' = \partial_{\tilde{\Omega}} \log \tilde{Z}', \quad 2\pi \tilde{M}' =  \left( - \partial_{\tilde{R}} + \frac{\tilde{\Omega}}{\tilde{R}} \partial_{\tilde{\Omega}}\right)\log \tilde{Z}', \quad \tilde{S}' =  \left(1 - \tilde{R} \partial_{\tilde{R}}\right) \log \tilde{Z}'.
\end{aligned}
\end{equation}
Note that here similar to $\tilde{Z}'$ we have also defined primed thermodynamic quantities which are simply themselves but without the dilaton dependence (for example, $\tilde{S} =e^{- \tilde{\Phi}_D} \tilde{S}'$). 
Now, combining the expression for $Q_n$ in (\ref{charges}) with (\ref{primerelations}), we get 
\begin{equation}\label{Qptrans}
\begin{aligned}
	2\pi R Q_n & = e^{-\Phi_D} \partial_{\mu_n}\left(\frac{R}{\tilde{R}} \log \tilde{Z}'   \right) = e^{-\Phi_D} R
	\left[ \frac{\partial \tilde{R}}{\partial \mu_n} \partial_{\tilde{R}} \left( \frac{1}{\tilde{R}} \log \tilde{Z}'\right) +   \frac{1}{\tilde{R}}
	\frac{\partial \tilde{\Omega}}{\partial \mu_n} \partial_{\tilde{\Omega}} \left( \log \tilde{Z}'\right)   \right] \\
	& = e^{-\Phi_D} R\left[ - \frac{1}{\tilde{R}^2}  \frac{\partial \tilde{R}}{\partial \mu_n}  \tilde{S}' + 2\pi \frac{\partial \tilde{\Omega}}{\partial \mu_n} \tilde{J}'
	  \right].
\end{aligned}
\end{equation}
Similar derivations lead to
\begin{equation}\label{Qwtrans}
	2\pi  Q_w = e^{-\Phi_D} \left[ - \frac{1}{\tilde{R}^2}  \frac{\partial \tilde{R}}{\partial \mu_w}  \tilde{S}' + 2\pi \frac{\partial \tilde{\Omega}}{\partial \mu_w} \tilde{J}'
	  \right] ,
\end{equation}
\begin{equation}\label{Jtrans}
\begin{aligned}
	2\pi J
	& = e^{-\Phi_D}  \left[ - \frac{1}{\tilde{R}^2} \frac{\partial \tilde{R} }{\partial  \Omega} \tilde{S}' + 2\pi  \frac{\partial \tilde{\Omega} }{\partial  \Omega}  \tilde{J}'\right],
\end{aligned}
\end{equation}
\begin{equation}\label{Strans}
\begin{aligned}
	S
	& = e^{-\Phi_D} \left[   \frac{R^2}{\tilde{R}^2} \frac{\partial \tilde{R}}{\partial R}  \tilde{S}' -  2\pi R^2 \frac{\partial \tilde{\Omega}}{\partial R} \tilde{J}'  \right],
\end{aligned}
\end{equation}
and finally, for the energy, we have
\begin{equation}\label{Etrans}
\begin{aligned}
	2\pi M & = 2\pi e^{-\Phi_D} \tilde{M}' + e^{-\Phi_D} \left[-1 + \frac{R}{\tilde{R}} \frac{\partial \tilde{R}}{\partial R} - \frac{\mu_n}{\tilde{R}}\frac{\partial \tilde{R}}{\partial \mu_n} - \frac{\mu_w}{\tilde{R}}\frac{\partial \tilde{R}}{\partial \mu_w}-\frac{\Omega}{\tilde{R}}\frac{\partial \tilde{R}}{\partial \Omega} \right] \frac{\tilde{S}'}{\tilde{R}} \\
	& \quad +2\pi e^{-\Phi_D} \left[ - 1 - \frac{R}{\tilde{\Omega}} \frac{\partial \tilde{\Omega}}{\partial R}+ \frac{\Omega}{\tilde{\Omega}} \frac{\partial \tilde{\Omega}}{\partial \Omega} +   \frac{\mu_n}{\tilde{\Omega}} \frac{\partial \tilde{\Omega}}{\partial \mu_n} + \frac{\mu_w}{\tilde{\Omega}} \frac{\partial \tilde{\Omega}}{\partial \mu_w}  \right] \tilde{\Omega}\tilde{J}'.
\end{aligned}
\end{equation}
(\ref{Qptrans}) - (\ref{Strans}) constitute as generalizations of the formulas in \cite{Chen:2021dsw}.
These relations are general and we didn't use the specific form of the solution generating transformation yet. All the information of the specific solution generation is contained in the map between
\begin{equation}
	\{\tilde{R}, \tilde{\Omega}\} \leftrightarrow \{ R, \Omega, \mu_n, \mu_w\}. 
\end{equation}
To find these relations, only the asymptotic form of the solution generating transformation is needed. To zeroth order in $\alpha'$, the transformation is the same in the bosonic/Type II and heterotic case, while they are different at order $\alpha'$, which we defer to \ref{sec:alphacorr}. Following the derivation of \cite{Chen:2021dsw} (which uses the formalism of \cite{Maharana:1992my}), we have\footnote{Here we follow the convention of \cite{Chen:2021dsw} for the normalization of the gauge fields and charges, which leads to a $\sqrt{2}$ factor difference compared to the main text, i.e. $Q_{\rm appendix} = Q_{\textrm{main text}}/\sqrt{2}$. }
\begin{equation}\label{RtildeR}
	\tilde{R} = R \sqrt{1-\mu_n^2}\sqrt{1-\mu_w^2}, \quad\quad \tilde{\Omega} = \frac{\Omega}{\sqrt{1-\mu_n^2}\sqrt{1-\mu_w^2}}.
\end{equation}
An easy way to understand the relation between $\Omega$ and $\tilde{\Omega}$ is the following. Let's say that the rotation is in $\varphi$ direction, then in the seed solution we have periodic identification $(t_{E} ,\varphi) \sim (t_{E} + 1, \varphi + 2\pi i\tilde{R} \tilde{\Omega})$, where we've chosen the Euclidean time coordinate such that its periodicity is one. This identification should remain invariant under the transformation, meaning that $\tilde{R}\tilde{\Omega}$ should be invariant, which explains (\ref{RtildeR}). To connect with the discussion in the main text, we can further parameterize $\mu_n$ and $\mu_w$ in terms of $\upalpha,\upbeta$
\begin{equation}\label{mu}
	\mu_n = \tanh \frac{\upalpha + \upbeta}{2}, \quad \quad \mu_{w} = \tanh \frac{\upalpha - \upbeta}{2}.
\end{equation}
Using (\ref{RtildeR}) and (\ref{mu}), as well as $Q_{L,R} = Q_n \pm Q_w$, we can simplify (\ref{Qptrans}) - (\ref{Strans}) and get
\begin{equation}\label{QLgene}
	2\pi Q_L = e^{-\Phi_D} \cosh \upbeta \sinh\upalpha \left[   \frac{\tilde{S}'}{\tilde{R}} + 2\pi \tilde{\Omega} \tilde{J}' \right],
\end{equation}
\begin{equation}
	2\pi Q_R =  e^{-\Phi_D} \cosh \upalpha  \sinh \upbeta \left[   \frac{\tilde{S}'}{\tilde{R}} + 2\pi \tilde{\Omega} \tilde{J}' \right],
\end{equation}
\begin{equation}
	2\pi J = 2\pi e^{-\Phi_D} \frac{\cosh \upalpha + \cosh \upbeta}{2} \tilde{J}',
\end{equation}
\begin{equation}
	S = e^{-\Phi_D} \tilde{S}' \frac{\cosh \upalpha + \cosh \upbeta}{2},
\end{equation}
\begin{equation}\label{Egene}
	2\pi M = 2\pi e^{-\Phi_D} \tilde{M}' + e^{-\Phi_D} (\cosh \upalpha \cosh\upbeta - 1) \left(\frac{\tilde{S}'}{\tilde{R}} +2\pi \tilde{\Omega} \tilde{J}'  \right).
\end{equation}
The point of (\ref{QLgene}) - (\ref{Egene}) is that once we know $\{ \tilde{M}', \tilde{S}', \tilde{J}'\}$ of the uncharged solution, 
the thermodynamic quantities of the charged solution will be completely determined. For example, a consistency check of these 
expressions is that if we plug in the $\{ \tilde{M}', \tilde{S}', \tilde{J}'\}$ for a four dimensional Kerr black 
hole, we correctly reproduce the thermodynamics of the two-charge black hole in \cite{Sen:1994eb}. 

As an aside, it is interesting to note from the expressions that the combination of $ \frac{\tilde{S}'}{\tilde{R}} + 2\pi \tilde{\Omega} \tilde{J}'$ is the ``charge" being transformed.

\subsection{Generating $\alpha'$ corrections in various cases}\label{sec:alphacorr}

A practical application of (\ref{QLgene}) - (\ref{Egene}) is that one could use them to work out the $\alpha'$ correction to the rotating two-charge system using the knowledge of the $\alpha'$ correction to the seed solution, without needing to find the explicit corrected solution. Of course, this is assuming that we don't encounter any singular solutions along the solution generating transformation, so perturbative $\alpha'$ corrections are under control. For this reason, one cannot apply the results here directly to the solution we studied in the main text, which is singular at the two derivative level. Nonetheless, the formulas would apply when we are far away from the singular limit $\Omega = 2\pi i /\beta$.  

The starting point of the derivation is the $\alpha'$ correction to the thermodynamics of the Kerr black hole (focusing on $D=4$). This can be dervied by simply evaluating the leading $\alpha'$ correction  to the action on the uncorrected black hole background. The end results, in the bosonic/heterotic cases, are
\begin{equation}\label{betaseed}
	\tilde{\beta} = \frac{4\pi r_+ (a^2+r_+^2)}{r_+^2 -a^2}, \quad \tilde{\Omega} = \frac{a}{r_+^2+a^2},
\end{equation}
\begin{equation}\label{Ekerr}
	\tilde{M}' = \frac{r_+^2 +a^2}{2r_+}  , \quad \tilde{S}' = \pi (r_+^2 + a^2) + 2\pi \lambda , \quad \tilde{J}' = a \tilde{M}'.
\end{equation}
Here $\lambda = \alpha'/2, \alpha'/4$ for bosonic and heterotic string theory, respectively. $\lambda$ vanishes for Type II theories and the leading order correction happens at $\alpha'^3$, therefore we won't discuss it here. We emphasize that $r_+,a$ in (\ref{betaseed}) and (\ref{Ekerr}) are only parameters to parametrize $\tilde{\beta}$ and $\tilde{\Omega}$ and they differ from the corresponding geometric quantities by $\alpha'$ corrections.

For bosonic string theory, we can then plug (\ref{betaseed}), (\ref{Ekerr}) into (\ref{QLgene}) - (\ref{Egene}) can get the leading $\alpha'$ corrections to the rotating two charge system:
\begin{equation}
	\frac{1}{T} = \frac{4\pi r_+ ( r_+^2 + a^2)}{r_+^2 -a^2} \frac{\cosh\upalpha + \cosh\upbeta}{2}, \quad \Omega = \frac{a}{r_+^2 +a^2} \frac{2}{\cosh\upalpha + \cosh \upbeta},
\end{equation}
\begin{equation}
	Q_L  = \sinh \upalpha \cosh \upbeta\left[ \frac{ \pi  (r_+^2 + a^2) }{2r_+ } +  \lambda \frac{ \pi  (r_+^2 - a^2) }{r_+ (r_+^2 + a^2)} \right],
\end{equation}
\begin{equation}
	Q_R  = \cosh \upalpha \sinh \upbeta\left[ \frac{ \pi  (r_+^2 + a^2) }{2r_+ } +  \lambda \frac{ \pi  (r_+^2 - a^2) }{r_+ (r_+^2 + a^2)} \right],
\end{equation}
\begin{equation}
	J = \frac{\cosh\upalpha + \cosh\upbeta}{2} a \frac{r_+^2 +a^2}{2r_+} ,
\end{equation}
\begin{equation}
	2\pi M = \frac{\pi (r_+^2 +a^2)}{2r_+} (1+ \cosh\upalpha \cosh \upbeta) + \lambda \frac{\pi (r_+^2 - a^2)}{r_+ (r_+^2 + a^2)} (\cosh \upalpha \cosh \upbeta - 1),
\end{equation}
\begin{equation}
	S = \frac{\cosh\upalpha + \cosh\upbeta}{2} (\pi (r_+^2 + a^2) + 2\pi \lambda ),
\end{equation}
where we've set $G_N = 1$ in the above expressions.

For the case of the heterotic string theory, the correct solution generating transformation in fact differs from the bosonic and type II theory by terms proportional to $\alpha'$. The correct transformation was motivated by studying the Horowitz-Polchinski solution in \cite{Chen:2021dsw}, but was later independently verified by studying the $\alpha'$ corrections to the non-rotating two-charge system \cite{Massai:2023cis}. The derivation here follows that of \cite{Chen:2021dsw}, with the main new ingredient being that the angular potential $\Omega$ is related to $\tilde{\Omega}$ of the seed by $R\Omega = \tilde{R}\tilde{\Omega}$. We omit the intermediate steps and present the results analogous to (\ref{QLgene}) to (\ref{Egene}). 
\begin{equation}\label{geneHet1}
    R = \frac{\tilde{R}}{2} \left[ \left(1 + \frac{\alpha'}{2\tilde{R}^2}\right)  \cosh \upalpha +  \left(1 - \frac{\alpha'}{2\tilde{R}^2}\right)  \cosh \upbeta \right], \quad \Omega = \frac{\tilde{R}}{R}\tilde{\Omega}, 
\end{equation}
\begin{equation}
    2\pi Q_L = e^{-\Phi_D} \cosh \upbeta \sinh\upalpha \left[   \frac{\tilde{S}'}{\tilde{R}} + 2\pi \tilde{\Omega} \tilde{J}' \right],\quad 2\pi Q_R = e^{-\Phi_D} \cosh \upalpha  \sinh \upbeta \left[   \frac{\tilde{S}'}{\tilde{R}} + 2\pi \tilde{\Omega} \tilde{J}' \right],
\end{equation}
\begin{equation}
     J = e^{-\Phi_D} \frac{\tilde{J}'}{2} \left[ \left(1 + \frac{\alpha'}{2\tilde{R}^2}\right)  \cosh \upalpha +  \left(1 - \frac{\alpha'}{2\tilde{R}^2}\right)  \cosh \upbeta \right], 
\end{equation}
\begin{equation}
    S = e^{-\Phi_D} \left[ \frac{\cosh \upalpha + \cosh \upbeta}{2} \tilde{S}' - \frac{\alpha' (\cosh \upalpha - \cosh\upbeta)}{4\tilde{R}^2} ( \tilde{S}' + 2\pi \tilde{R} \tilde{\Omega} \tilde{J}')\right],
\end{equation}
\begin{equation}\label{geneHet5}
   2\pi M =  2\pi e^{-\Phi_D} \tilde{M}' + e^{-\Phi_D} (\cosh \upalpha \cosh\upbeta - 1) \left(\frac{\tilde{S}'}{\tilde{R}} +2\pi \tilde{\Omega} \tilde{J}'  \right).
\end{equation}
We can further plug (\ref{betaseed}) and (\ref{Ekerr}) and get the explicit formulas for the $\alpha'$ corrected thermodynamic quantities for rotating two-charge solutions in heterotic string theory.

\bibliography{SmallBHIndex}
\bibliographystyle{JHEP}

\end{document}